\documentclass[final,5p,times,twocolumn]{elsarticle}

\usepackage{amssymb}
\usepackage{hyperref}
\usepackage{braket}
\usepackage{graphicx}
\usepackage{pdflscape}
\usepackage{afterpage}

 \biboptions{sort&compress}

\newcounter{bla}

\journal{Computer Physics Communications}

\begin{document}

\begin{frontmatter}



\title{CheMPS2: a free open-source spin-adapted implementation of the density matrix renormalization group for ab initio quantum chemistry}


\author[a]{Sebastian {Wouters}\corref{author}}
\author[a]{Ward {Poelmans}}
\author[b]{Paul W. {Ayers}}
\author[a]{Dimitri {Van Neck}}

\cortext[author] {Corresponding author.\\\textit{E-mail address:} sebastianwouters@gmail.com\\\textit{Phone number:} +32 9 264 6641}
\address[a]{Center for Molecular Modeling, Ghent University, Technologiepark 903, 9052 Zwijnaarde, Belgium}
\address[b]{Department of Chemistry, Mc{M}aster University, Hamilton, Ontario L8S 4M1, Canada}

\begin{abstract}
The density matrix renormalization group (DMRG) has become an indispensable numerical tool to find exact eigenstates of finite-size quantum systems with strong correlation. In the fields of condensed matter, nuclear structure and molecular electronic structure, it has significantly extended the system sizes that can be handled compared to full configuration interaction, without losing numerical accuracy. For quantum chemistry (QC), the most efficient implementations of DMRG require the incorporation of particle number, spin and point group symmetries in the underlying matrix product state (MPS) ansatz, as well as the use of so-called complementary operators. The symmetries introduce a sparse block structure in the MPS ansatz and in the intermediary contracted tensors. If a symmetry is non-abelian, the Wigner-Eckart theorem allows to factorize a tensor into a Clebsch-Gordan coefficient and a reduced tensor. In addition, the fermion signs have to be carefully tracked. Because of these challenges, implementing DMRG efficiently for QC is not straightforward. Efficient and freely available implementations are therefore highly desired. In this work we present CheMPS2, our free open-source spin-adapted implementation of DMRG for ab initio QC. Around CheMPS2, we have implemented the augmented Hessian Newton-Raphson complete active space self-consistent field method, with exact Hessian. The bond dissociation curves of the 12 lowest states of the carbon dimer were obtained at the DMRG(28 orbitals, 12 electrons, D$_{ \mathsf{ SU(2) }}$=2500)/cc-pVDZ level of theory. The contribution of $1s$ core correlation to the $X^1\Sigma_g^+$ bond dissociation curve of the carbon dimer was estimated by comparing energies at the DMRG(36o, 12e, D$_{\mathsf{SU(2)}}$=2500)/cc-pCVDZ and DMRG-SCF(34o, 8e, D$_{\mathsf{SU(2)}}$=2500)/cc-pCVDZ levels of theory.

\end{abstract}

\begin{keyword}
density matrix renormalization group \sep matrix product state \sep SU(2) spin symmetry \sep abelian point group symmetry \sep ab initio quantum chemistry

\end{keyword}

\end{frontmatter}



{\bf PROGRAM SUMMARY}

\begin{small}
\noindent
{\em Manuscript Title:} CheMPS2: a free open-source spin-adapted implementation of the density matrix renormalization group for ab initio quantum chemistry                         \\
{\em Authors:} Sebastian Wouters, Ward Poelmans, Paul W. Ayers, and Dimitri {Van Neck} \\
{\em Program Title:} CheMPS2                                  \\
{\em Journal Reference:}                                      \\
{\em Catalogue identifier:}                                   \\
{\em Licensing provisions:} GNU General Public License version 2 \\
{\em Programming language:} C++                               \\
{\em Computer:} x86-64                                        \\
{\em Operating system:} Scientific Linux 6.0                  \\
{\em RAM:} 10 MB - 64 GB                                        \\
{\em Number of processors used:} 1 - 16 (single node)         \\
{\em Supplementary material:} Doxygen documentation can be generated \\
{\em Keywords:} density matrix renormalization group, matrix product state, SU(2) spin symmetry, abelian point group symmetry, ab initio quantum chemistry                         \\
{\em Classification:} 16.1 Molecular Physics and Physical Chemistry: Structure and Properties \\
{\em External routines/libraries:} Basic Linear Algebra Subprograms (BLAS), Linear Algebra Package (LAPACK), GNU Scientific Library (GSL), and Hierarchical Data Format Release 5 (HDF5) \\
{\em Nature of problem:}\\
The many-body Hilbert space grows exponentially with the number of single-particle states. Exact diagonalization solvers can therefore only handle small systems, of up to 18 electrons in 18 orbitals. Interesting active spaces are often significantly larger.
   \\
{\em Solution method:}\\
The density matrix renormalization group allows to extend the size of active spaces, for which numerically exact solutions can be found, to about 40 electrons in 40 orbitals. In addition, it provides a rigorous variational upper bound to energies, as it has an underlying wavefunction ansatz, the matrix product state.
   \\
{\em Restrictions:}\\
Our implementation of the density matrix renormalization group is spin-adapted. This means that targeted eigenstates in the active space are exact eigenstates of the total electronic spin operator. Hamiltonians which break this symmetry (a magnetic field term for example) cannot be handled by our code. As electron repulsion integrals in gaussian basis sets have eightfold permutation symmetry, we have used this property in our code.
   \\
{\em Unusual features:}\\
The nature of the matrix product state ansatz allows for exact spin coupling. In CheMPS2, the \textit{total} electronic spin is imposed (not just the spin projection), in addition to the particle-number and abelian point-group symmetries.
   \\
{\em Running time:}\\
The running time depends on the size of the targeted active space, the number of desired eigenstates, their symmetry, the density of states, the individual orbital symmetries, the orbital ordering, the desired level of convergence, and the chosen convergence scheme. To converge a single point of one of the dissociation curves of the carbon dimer ($D_{\infty h} \rightarrow D_{2h}$ symmetry) in the cc-pVDZ basis (28 orbitals; their ordering is described in section \ref{secIrrepOrder}) with 2500 reduced renormalized basis states (see the convergence scheme in section \ref{HowToExtrapolateEnergies}; the variational energy then lies 0.1 $mE_h$ above the fully converged result) takes about 8 hours on a single node with a dual-socket octa-core Intel Xeon Sandy Bridge (E5-2670) (16 cores at 2.6 GHz), and requires 6 GB of RAM.
   \\

\end{small}

\section{Introduction} \label{intro}
Conventional molecular electronic structure methods such as density functional theory, Hartree-Fock theory, and coupled cluster theory start with the assumption that a single Slater determinant (SD) provides a qualitatively good description of the molecule at hand \cite{BookHelgaker}. While this assumption is valid for some molecules near equilibrium geometry, the static correlation which arises in other molecules, as well as for geometries far from equilibrium, requires the use of multireference (MR) methods. These provide a qualitative description which is equivalent to multiple SDs, thereby resolving the static correlation. One of these MR methods is the exact diagonalization of the many-body Hamiltonian in the full Hilbert space, also known as full configuration interaction (FCI) in quantum chemistry (QC). Because the many-body Hilbert space grows exponentially with the number of single-particle states, only small systems, of up to 18 electrons in 18 orbitals, can be treated by FCI. In 1999, the density matrix renormalization group (DMRG) was introduced in QC \cite{whiteQC}. This MR method allows to extend the system sizes for which numerically exact solutions can be found to about 40 electrons in 40 orbitals, depending on the nature of the system.

DMRG originated in 1992 in the field of condensed matter \cite{PhysRevLett.69.2863, PhysRevB.48.10345}. Although it was originally introduced as a renormalization group flow for increasing many-body Hilbert spaces, in 1995 it was realized that DMRG can be reformulated as the variational optimization of a particular wavefunction ansatz, the matrix product state (MPS) \cite{PhysRevLett.75.3537,PhysRevB.55.2164}. This not only provided the theoretical validation that an energy obtained with DMRG is always an upper bound to the exact eigenvalue, but also shed light on DMRG from a quantum information perspective. Non-critical quantum mechanical ground states are believed to obey the so-called area law for the entanglement entropy \cite{1742-5468-2007-08-P08024}. This implies that quantum correlation is local in such a ground state. For one-dimensional systems, the boundary of a line segment consists of two points, and the entanglement entropy is a constant, independent of system length. This is the reason why DMRG works extremely well for one-dimensional non-critical systems. Quantum information theory also induced the development of other so-called tensor network states (TNS), which capture the entanglement entropy well in higher dimensional and/or critical systems \cite{PEPSverstraete,PhysRevLett.99.220405}. There even exists a continuous MPS ansatz for quantum fields \cite{PhysRevLett.104.190405}.

Although the active orbital space of most molecular systems is far from one-dimensional, DMRG has been very useful for ab initio QC \cite{whiteQC, QUA:QUA1, mitrushenkov:6815, chan:4462, PhysRevB.67.125114, chan:8551, doi:10.1080/0026897031000155625, mitrushenkov:4148, PhysRevB.68.195116, chan:3172, chan:6110, PhysRevB.70.205118, moritz:024107, chan:204101, moritz:184105, moritz:034103, hachmann:144101, Rissler2006519, moritz:244109, dorando:084109, hachmann:134309, marti:014104, zgid:014107, zgid:144115, zgid:144116, ghosh:144117, ChanB805292C, ChanQUA:QUA22099, dorando:184111,kurashige:234114,yanai:024105, neuscamman:024106,doi:10.1080/00268971003657078, PhysRevB.81.235129, mizukami:091101, PhysRevA.83.012508, boguslawski:224101, kurashige:094104, QUA:QUA23173, sharma:124121, wouters, doi:10.1021/ct300211j, C2CP23767A, doi:10.1021/jz301319v, doi:10.1021/ct3008974, doi:10.1021/ct400247p, naturechem, ma:224105, saitow:044118, doi:10.1021/ct400707k, C3CP53975J, KnechtPaper}. The variational upper bound to the true eigenvalue, obtained with DMRG, can be systematically improved by increasing the so-called bond or virtual dimension of the MPS ansatz. This provides a way to check the convergence of DMRG calculations.

In ab initio QC methods which use FCI, the FCI solver can be replaced by DMRG. Ab initio DMRG allows for an efficient extraction of the reduced two-body density matrix (2-RDM) \cite{zgid:144115}. The 2-RDM of the active space is required in the complete active space self-consistent field (CASSCF) method to compute the gradient and the Hessian. It is therefore natural to introduce a CASSCF variant with DMRG as active space solver, DMRG-SCF \cite{zgid:144116}. This allows one to describe static correlation in large active spaces. To add dynamic correlation as well, three DMRG-based methods have been introduced. \textit{(a)} With a little more effort, the 3-RDM and contracted 4-RDMs can be extracted from DMRG as well. These are required to apply second order perturbation theory to a CASSCF wavefunction, called CASPT2. The DMRG variant is DMRG-CASPT2 \cite{kurashige:094104}. \textit{(b)} Based on a CASSCF wavefunction, a configuration interaction expansion can be introduced, called MRCI. Recently, an approximate DMRG-MRCI variant was proposed \cite{saitow:044118}. \textit{(c)} Yet another way is to perform a canonical transformation (CT) on top of an MR wavefunction. When an MPS is used as MR wavefunction, the method is called DMRG-CT \cite{yanai:024105}.

In addition to ground states, DMRG can also find excited states. By projecting out lower lying eigenstates, or by targeting a specific energy \cite{dorando:084109}, the DMRG algorithm solves for a particular excited state. In these state-specific algorithms, the whole renormalized basis is used to represent one single eigenstate. In state-averaged DMRG, several eigenstates are targeted at once. Their RDMs are weighted and summed to perform the DMRG renormalization step \cite{HallbergBook}. The renormalized basis then represents several eigenstates at once.

DMRG linear response theory (DMRG-LRT) can be used as well to find excited states. Once the ground state has been found, the MPS tangent vectors to this optimized point can be used as an (incomplete) variational basis to approximate excited states \cite{dorando:184111, PhysRevB.85.035130, PhysRevB.85.100408, PhysRevB.88.075122, PhysRevB.88.075133, 2013arXiv1311.1646N}. As the tangent vectors to an optimized SD yield the configuration interaction with singles (CIS), also called the Tamm-Dancoff approximation (TDA), for Hartree-Fock theory \cite{BookHelgaker}, the same names are used for DMRG: DMRG-CIS or DMRG-TDA. By linearizing the time-dependent variational principle for matrix product states \cite{PhysRevLett.107.070601}, the DMRG random phase approximation (DMRG-RPA) is found \cite{PhysRevB.88.075122, PhysRevB.88.075133, 2013arXiv1311.1646N}, again in complete analogy with RPA for Hartree-Fock theory. The variational optimization in an (incomplete) basis of MPS tangent vectors can be extended to higher-order tangent spaces as well. DMRG-CISD, or DMRG configuration interaction with singles and doubles, is a variational approximation to target both ground and excited states in the space spanned by the MPS reference and its single and double tangent spaces \cite{PhysRevB.88.075122}.

In ab initio QC, two other TNSs have been employed as well: the tree TNS \cite{PhysRevB.82.205105, nakatani:134113} and the complete-graph TNS \cite{1367-2630-12-10-103008}. While they require a smaller virtual dimension to achieve the same accuracy, their optimization algorithms are less efficient, and as a result an MPS is currently still the preferred choice for ab initio QC. 

In section \ref{remarks}, the DMRG algorithm is briefly introduced, and remarks specific to ab initio QC are discussed. In section \ref{symm}, the implementation of particle number, spin, and abelian point group symmetries is presented. An overview of the structure of CheMPS2 is given in section \ref{ourcode}. Results on the low-lying states of the carbon dimer are presented in section \ref{C2}.
A summary is given in section \ref{summary}. Atomic units are used in this work: $E_h = 4.35974434(19) \times 10^{-18}$ J and $a_0 = 5.2917721092(17) \times 10^{-11}$ m \cite{RevModPhys.84.1527}.

\section{DMRG for ab initio quantum chemistry} \label{remarks}

\subsection{The MPS ansatz}
DMRG can be formulated as the variational optimization of an MPS. The MPS ansatz with open boundary conditions is given by
\begin{eqnarray}
\ket{\Psi} & = & \sum\limits_{\{ n_k \}, \{ \alpha_j \}} A[1]^{n_1}_{\alpha_1} A[2]^{n_2}_{\alpha_1;\alpha_2} ... A[L-1]^{n_{L-1}}_{\alpha_{L-2};\alpha_{L-1}} A[L]^{n_L}_{\alpha_{L-1}} \nonumber \\
           &   & ~ \ket{n_1 n_2 ... n_L}
\end{eqnarray}
where $n_k$ denotes the occupancy of orbital $k$ ($\ket{-}$, $\ket{\uparrow}$, $\ket{\downarrow}$, or $\ket{\uparrow\downarrow}$) and the $\{\alpha_j\}$ are the so-called bond or virtual indices. With increasing dimension $D$ of these virtual indices, a larger part of the Hilbert space can be reached. Note that it is of no use to make virtual dimension $D_j$ larger than min$(4^j,4^{L-j})$, the minimum of the sizes of the partial Hilbert spaces spanned by resp. the first $j$ and the last $L-j$ orbitals.

\subsection{Canonical forms}
The wavefunction $\ket{\Psi}$ does not uniquely define the ansatz, in analogy with a Slater determinant. For the latter, a rotation in the occupied orbital space alone, or a rotation in the virtual orbital space alone, does not change the physical wavefunction. Only occupied-virtual rotations change the wavefunction. In an MPS, there is gauge freedom as well. If for two neighbouring sites $i$ and $i+1$, the left MPS tensors are right-multiplied with the non-singular matrix $G$
\begin{equation}
\tilde{A}[i]^{n_i}_{\alpha_{i-1};\alpha_i} = \sum\limits_{\alpha_j} A[i]^{n_i}_{\alpha_{i-1};\alpha_j} G_{\alpha_j;\alpha_i}
\end{equation}
and the right MPS tensors are left-multiplied with the inverse of $G$
\begin{equation}
\tilde{A}[i+1]^{n_{i+1}}_{\alpha_{i};\alpha_{i+1}} = \sum\limits_{\alpha_j} G^{-1}_{\alpha_{i};\alpha_{j}} A[i+1]^{n_{i+1}}_{\alpha_{j};\alpha_{i+1}}
\end{equation}
the wavefunction does not change, i.e. $\forall n_i,n_{i+1},\alpha_{i-1},\alpha_{i+1}$:
\begin{equation}
\sum\limits_{\alpha_i} \tilde{A}[i]^{n_i}_{\alpha_{i-1};\alpha_i} \tilde{A}[i+1]^{n_{i+1}}_{\alpha_{i};\alpha_{i+1}} = \sum\limits_{\alpha_i} A[i]^{n_i}_{\alpha_{i-1};\alpha_i} A[i+1]^{n_{i+1}}_{\alpha_{i};\alpha_{i+1}}.
\end{equation}
CheMPS2 is a two-site DMRG algorithm, were at each so-called micro-iteration two neighbouring sites are simultaneously optimized. Suppose these sites are $i$ and $i+1$. The gauge freedom of the MPS is used to bring it in a particular canonical form. For all sites to the left of $i$, the MPS tensors are left-normalized:
\begin{equation}
\sum\limits_{\alpha_{k-1}, n_k} \left(A[k]^{n_k}\right)^{\dagger}_{\alpha_k; \alpha_{k-1}} A[k]^{n_k}_{\alpha_{k-1};\beta_k} = \delta_{\alpha_k, \beta_k} \label{left-normalized}
\end{equation}
and for all sites to the right of $i+1$, the MPS tensors are right-normalized:
\begin{equation}
\sum\limits_{\alpha_{k}, n_k} A[k]^{n_k}_{\alpha_{k-1};\alpha_k} \left(A[k]^{n_k}\right)^{\dagger}_{\alpha_k; \beta_{k-1}} = \delta_{\alpha_{k-1}, \beta_{k-1}}. \label{right-normalized}
\end{equation}

\subsection{The effective Hamiltonian equation} \label{secEffHam}
Combine the MPS tensors of the two sites under consideration into a single two-site tensor:
\begin{equation}
\sum\limits_{\alpha_i} A[i]^{n_i}_{\alpha_{i-1};\alpha_i} A[i+1]^{n_{i+1}}_{\alpha_{i};\alpha_{i+1}} = B[i]_{\alpha_{i-1};\alpha_{i+1}}^{n_i;n_{i+1}}. \label{two-site-object-not-reduced}
\end{equation}
At the current micro-iteration of the DMRG algorithm, $\mathbf{B}[i]$ (the flattened form of the tensor $B[i]$) is used as an initial guess for the effective Hamiltonian equation. This equation is obtained by variation of the Lagrangian \cite{ChanB805292C}
\begin{equation}
\mathcal{L} = \braket{\Psi(\mathbf{B}[i]) \mid \hat{H} \mid \Psi(\mathbf{B}[i])} - \lambda \braket{\Psi(\mathbf{B}[i]) \mid \Psi(\mathbf{B}[i])}
\end{equation}
to the complex conjugate of $\mathbf{B}[i]$:
\begin{equation}
\mathbf{H}^{eff} \mathbf{B}[i] = \lambda \mathbf{B}[i]. \label{effHameq}
\end{equation}
The specific canonical choice of Eqs. (\ref{left-normalized})-(\ref{right-normalized}) ensured that no overlap matrix is present in this effective Hamiltonian equation. The lowest eigenvalue and corresponding eigenvector of this equation are searched. In CheMPS2, this is done with our implementation of Davidson's algorithm \cite{Davidson197587}. Once found, it is decomposed with a singular value decomposition:
\begin{equation}
B[i]_{ \left( \alpha_{i-1} n_i \right) ; \left( n_{i+1} \alpha_{i+1} \right)} = \sum\limits_{\beta} U[i]_{ \left( \alpha_{i-1} n_i \right) ; \beta} \kappa[i]_{\beta} V[i]_{ \beta ; \left( n_{i+1} \alpha_{i+1} \right)}
\end{equation}
Note that $U[i]$ is hence left-normalized and $V[i]$ right-normalized. In the DMRG algorithm, the original sum over $\beta$ of dimension $\min(4D_{i-1},4D_{i+1})$ is truncated to $D_i$, thereby keeping the $D_i$ largest $\kappa[i]_{\beta}$.

\subsection{Sweeping}
So far, we have looked at a micro-iteration of the DMRG algorithm. This micro-iteration happens during left or right sweeps. During a left sweep, $B[i]$ is constructed, the corresponding effective Hamiltonian equation solved, the solution $B[i]$ decomposed, the singular value spectrum truncated, $A[i]$ is set to $U[i] \times \kappa[i]$, $A[i+1]$ is set to $V[i]$, and $i$ is decreased by 1. Note that $A[i+1]$ is right-normalized for the next micro-iteration as required. This stepping to the left occurs until $i=0$, and then the sweep direction is reversed from left to right. Based on energy differences, or wavefunction overlaps, between consecutive sweeps, a convergence criterium is triggered, and the sweeping stops. One sweep is called a macro-iteration in DMRG.

\subsection{Complementary operators}
The effective Hamiltonian in Eq. (\ref{effHameq}) is too large to be fully constructed. Only its action on a particular guess $\mathbf{B}[i]$ is available as a function. In order to construct $\mathbf{H}^{eff} \mathbf{B}[i]$ efficiently for general quantum chemistry Hamiltonians, several tricks are used. \textit{(a)} The one-body matrix elements $(i|T|k)$ are incorporated in the two-body matrix elements $(ij|V|kl)$:
\begin{equation}
(ij|h|kl) = (ij|V|kl) + \frac{1}{N-1} \left[ (i|T|k) \delta_{j,l} + (j|T|l) \delta_{i,k} \right] 
\end{equation}
where $N$ is the targeted particle number. \textit{(b)} Suppose we want to optimize sites $i$ and $i+1$, and that $\ket{\alpha_{i-1}}$ are the corresponding $D_{i-1}$ left renormalized basis states. Renormalized operators such as $\braket{\alpha_{i-1} \mid \hat{a}_{k\sigma}^{\dagger} \hat{a}_{l\tau} \mid \beta_{i-1}}$ with $k$ and $l$ both smaller than $i$ are constructed and stored on disk \cite{chan:4462}. For the second quantized operators $\hat{a}^{\dagger}$ and $\hat{a}$, the Latin indices denote orbitals and the Greek indices spin projections. \textit{(c)} Once three second quantized operators are on one side of $B[i]$, they are multiplied with the matrix elements $(ij|h|kl)$, and a summation is performed over the common indices to construct complementary operators \cite{PhysRevB.53.R10445}:
\begin{eqnarray}
\sum\limits_{\sigma} \sum\limits_{k,l,m<i} \braket{\alpha_{i-1} \mid \hat{a}_{k\sigma}^{\dagger} \hat{a}^{\dagger}_{l\tau} \hat{a}_{m\sigma} \mid \beta_{i-1}} \times (kl|h|mn) & \nonumber \\
 \rightarrow \braket{\alpha_{i-1} \mid \hat{O}_{n\tau} \mid \beta_{i-1}}. &
\end{eqnarray}
For two, three, and four second quantized operators on one side of $B[i]$, these complementary operators are constructed. A bare (without matrix elements) renormalized operator is only constructed for one or two second quantized operators on one side of $B[i]$. \textit{(d)} Hermitian conjugation
\begin{equation}
\braket{\alpha_{i-1} \mid \hat{a}_{k\sigma}^{\dagger} \hat{a}^{\dagger}_{l\tau} \mid \beta_{i-1}} = \braket{\beta_{i-1} \mid \hat{a}_{l\tau} \hat{a}_{k\sigma} \mid \alpha_{i-1}}^{\dagger}
\end{equation}
and commutation relations between the second quantized operators are also used to further limit the storage requirement for the renormalized partial Hamiltonian terms.

\subsection{Convergence} \label{ConvergenceSection}
There is also a one-site DMRG algorithm, in which only one MPS site tensor is optimized at each micro-iteration, but this algorithm is more likely to get stuck in a local minimum. To help prevent the two-site DMRG algorithm from getting stuck in a local minimum, a small amount of noise can be added to the solution $B[i]$, just before it is decomposed. This way, renormalized basis states corresponding to lost symmetries (which should be there, but are not) can be reintroduced \cite{chan:4462}.

The choice of orbitals and their ordering on the one-dimensional DMRG lattice have a significant influence both on getting stuck in local minima, as well as on how fast the variational energy $E_D$ converges with increasing $D$ \cite{whiteQC}. The optimal choice and ordering are still under debate, although two rules of thumb are widely used. Active space orbitals in elongated molcular systems (think about polyenes for example) should be localized as much as possible to respect the area law for the entanglement entropy \cite{QUA:QUA23173}. For small molecules with a high point group symmetry, it is beneficial to put bonding and anti-bonding orbitals close to each other on the one-dimensional DMRG lattice, as they are most strongly correlated \cite{ma:224105}.

One possibility to settle this ongoing debate might be to look at the so-called two-orbital mutual information $I_{p,q}$ in the future \cite{Rissler2006519}. This is a measure from quantum information theory for the amount of correlation between two orbitals, and is a two-point correlation function on the one-dimensional DMRG lattice. A cost function can be associated with this measure, e.g. $F = \sum_{p,q} I_{p,q} (p-q)^z$, which requires highly correlated orbitals to be close. Its gradient and Hessian with respect to orbital rotations can be calculated by resp. three- and four-point correlation functions on the one-dimensional DMRG lattice. These can be obtained efficiently \cite{zgid:144115}. If local minima can be avoided, this yields a set of minimally entangled orbitals and their optimal ordering, from which extra rules of thumb can be drawn.

Two extrapolation schemes exist to assess the convergence of the variational energy $E_D$ with increasing number of renormalized basis states $D$. The first is the scaling relation
\begin{equation}
\ln(E_D - E_{exact} ) = C_1 - C_2 (\ln(D))^2 
\end{equation}
proposed by Chan \cite{chan:4462,chanExtraPolWithAyers,wouters} which is nowadays not often used. The $C_i$ are constants which are determined by the fit. The second and most widely used extrapolation scheme is based on the so-called maximal discarded weight $w^{disc}(D)$ during the last DMRG sweep for a certain value of $D$:
\begin{equation}
w^{disc}(D) = \max\limits_{i} \left\{ \sum\limits_{\beta = D+1}^{4D} \kappa[i]^2_{\beta} \right\}.
\end{equation}
It proposes a linear relation between the variational energy $E_D$ and the discarded weight $w^{disc}(D)$ \cite{PhysRevB.53.14349,chan:4462, 2013arXiv1307.1002V}:
\begin{equation}
E(D) = E_{exact} + C_1 ~ w^{disc}(D). \label{extrapolSchemeEq}
\end{equation}
By increasing $D$ stepwise, $E_{exact}$ can be extrapolated.

\section{Symmetry-adapted DMRG} \label{symm}

\subsection{Introduction} \label{symm-intro}

The symmetry group of the Hamiltonian can be used to label eigenstates by symmetry. To find an eigenstate with a particular symmetry, it is sufficient to restrict an optimization to the corresponding corner of the many-body Hilbert space. For DMRG, it is well understood how both abelian and non-abelian symmetries can be imposed \cite{2002EL57852M,2007JSMTE1014M,2010NJPh12c3029S,PhysRevA.82.050301}. Each MPS tensor and intermediary contracted tensor decompose into a Clebsch-Gordan coefficient and a reduced tensor. The Clebsch-Gordan coefficient introduces a sparse block structure in the reduced tensor. If the symmetry group of the Hamiltonian is non-abelian, some irreducible representations (irrep) have a dimension larger than one, and then this factorization also presents an information compression, as the size of the full tensor is larger than the size of the reduced tensor. In addition to the possibility of restricting an optimization to a particular symmetry corner of the many-body Hilbert space, this sparsity and compression result in smaller requirements in disk, memory and computer time.

In CheMPS2, we have implemented three global symmetries for the MPS wavefunction: $\mathsf{SU(2)}$ total electronic spin, $\mathsf{U(1)}$ particle number, and abelian point group symmetry $\mathsf{P}$. As we work real-valued in CheMPS2, the latter are restricted to $\mathsf{P} \in \left\{C_1, C_i, C_2, C_s, D_2, C_{2v}, C_{2h}, D_{2h} \right\}$ \cite{BookCornwell}.

\subsection{Reduced MPS tensors}

These global symmetries are imposed by requiring that the MPS site tensors $A[i]^{n_i}_{\alpha_{i-1};\alpha_i}$ are irreducible tensor operators of the total symmetry group \cite{PhysRevA.82.050301, 2010NJPh12c3029S, 2007JSMTE1014M, 2002EL57852M}. The local and virtual basis states ($\ket{n_k}$ and $\ket{\alpha_j}$) then have to transform according to the rows of the irreps of this symmetry group. This is realized by rotating the basis states so that they can be represented by good spin ($s$ and $j$), spin projection ($s^z$ and $j^z$), particle number ($N$), and point group irrep ($I$) quantum numbers.

The local basis states of orbital $k$ are labeled as
\begin{eqnarray}
\ket{-} & \rightarrow & \ket{s=0; s^z=0, N=0; I=I_0} \\
\ket{\uparrow} & \rightarrow & \ket{s=\frac{1}{2}; s^z=\frac{1}{2}, N=1; I=I_k} \\
\ket{\downarrow} & \rightarrow & \ket{s=\frac{1}{2}; s^z=-\frac{1}{2}, N=1; I=I_k} \\
\ket{\uparrow\downarrow} & \rightarrow & \ket{s=0; s^z=0, N=2; I=I_0}
\end{eqnarray}
where $I_0$ and $I_k$ are resp. the trivial and orbital $k$ point group irreps. $\ket{\uparrow\downarrow}$ corresponds to $I_0$ because for the abelian point groups with real-valued character tables, $\forall I_k : I_k \otimes I_k = I_0$. In the same way, the virtual basis states are labeled as
\begin{equation}
\ket{\alpha} \rightarrow \ket{j j^z N I \alpha}
\end{equation}
where the $\alpha$ on the right-hand side allows to distinguish between seperate virtual basis states which belong to the same symmetry.

Due to the Wigner-Eckart theorem, each irreducible tensor operator $A[i]$ factorizes into Clebsch-Gordan coefficients and a reduced tensor $T[i]$:
\begin{eqnarray}
& A[i]^{n_i}_{\alpha_{i-1};\alpha_i} = A[i]^{s s^z N I}_{j_L j_L^z N_L I_L \alpha_{i-1}; j_R j_R^z N_R I_R \alpha_i} \nonumber\\
& = \braket{j_L j_L^z s s^z | j_R j_R^z} \delta_{N_L + N, N_R} \delta_{I_L \otimes I, I_R} T[i]^{(s N I)}_{(j_L N_L I_L \alpha_L)(j_R N_R I_R \alpha_R)} \quad \label{tensordecomp}
\end{eqnarray}
The $\mathsf{SU(2)}$, $\mathsf{U(1)}$, and $\mathsf{P}$ symmetries are imposed by their corresponding Clebsch-Gordan coefficients, and express nothing else than resp. local allowed spin recoupling, local particle conservation, and local point group symmetry conservation. The indices $\alpha_L$ and $\alpha_R$ keep track of the number of times an irrep occurs at a virtual bond. If the virtual dimension of a symmetry sector is $D(j_L N_L I_L)$, this would correspond to a dimension of $(2 j_L + 1) D(j_L N_L I_L)$ in an MPS which is not symmetry-adapted \cite{2002EL57852M}. If a Clebsch-Gordan coefficient is zero by symmetry, the corresponding blocks in $T[i]$ do not need to be allocated, resulting in sparse block structure. If $j$ or $s$ are not spin-0, there is in addition data compression.

The desired global symmetry can be imposed on the MPS by requiring that the left virtual index of the leftmost tensor in the MPS chain consists of one irrep corresponding to $(j_L, N_L, I_L) = (0,0,I_0)$, while the right virtual index of the rightmost tensor consists of one irrep corresponding to $(j_R, N_R, I_R) = (S_G, N_G, I_G)$, the desired global spin, particle number, and point group symmetry. This corresponds to the singlet-embedding strategy of Sharma and Chan \cite{sharma:124121}.

The operators
\begin{eqnarray}
\hat{b}^{\dagger}_{k \sigma} & = & \hat{a}^{\dagger}_{k \sigma} \label{creaannih1}\\
\hat{b}_{k \sigma} & = & (-1)^{\frac{1}{2}-\sigma}\hat{a}_{k -\sigma} \label{creaannih2}
\end{eqnarray}
for orbital $k$ correspond to resp. the $(s=\frac{1}{2}, s^z=\sigma, N=1, I_k)$ row of irrep $(s=\frac{1}{2}, N=1, I_k)$ and the $(s=\frac{1}{2}, s^z=\sigma, N=-1, I_k)$ row of irrep $(s=\frac{1}{2}, N=-1, I_k)$ \cite{BookDimitri}. $\hat{b}^{\dagger}$ and $\hat{b}$ are hence both doublet irreducible tensor operators. This fact permits exploitation of the Wigner-Eckart theorem also for renormalized operators and complementary operators, and to develop a code without any spin projections or $\mathsf{SU(2)}$ Clebsch-Gordan coefficients. Contracting terms of the type (\ref{tensordecomp}) and (\ref{creaannih1})-(\ref{creaannih2}) can be done by implicitly summing over the common multiplets and recoupling the local, virtual and operator spins. An example is given in \ref{redtensors}. Operators and complementary operators then formally consist of terms containing Clebsch-Gordan coefficients and reduced tensors. In our code, however, only the reduced tensors need to be calculated and stored. CheMPS2 uses the GNU Scientific Library \cite{GSLcitation} to extract Wigner 6-j and 9-j symbols for the recoupling. No Wigner 3-j symbols or Clebsch-Gordan coefficients are used in the program.

\subsection{The reduced two-site object}

Section \ref{secEffHam} can be reformulated with the reduced $T$-tensors from Eq. (\ref{tensordecomp}) and a reduced two-site object $S[i]$:
\begin{eqnarray}
& S[i]^{j(s_1 s_2) N_1 N_2 I_1 I_2}_{j_L N_L I_L \alpha_L ; j_R N_R I_R \alpha_R} = \delta_{N_L+N_1+N_2,N_R} \delta_{I_L \otimes I_1 \otimes I_2, I_R} \sqrt{2j+1} \nonumber \\
& (-1)^{j_L+j_R+s_1+s_2} \sum\limits_{j_M \alpha_M} \sqrt{2j_M+1} \left\{ \begin{array}{ccc} j_L & j_R & j \\ s_2 & s_1 & j_M \end{array} \right\} \nonumber\\
& T[i]^{s_1 N_1 I_1}_{j_L N_L I_L \alpha_L ; j_M (N_L+N_1) (I_L \otimes I_1) \alpha_M} \nonumber \\
& T[i+1]^{s_2 N_2 I_2}_{j_M (N_L+N_1) (I_L \otimes I_1) \alpha_M ; j_R N_R I_R \alpha_R}. \label{TTtoS}
\end{eqnarray}
Eq. (\ref{TTtoS}) is the analogue of Eq. (\ref{two-site-object-not-reduced}). The Lagrangian can be written in terms of $S[i]$, the effective Hamiltonian equation can be solved, and after convergence, Eq. (\ref{TTtoS}) can be backtransformed:
\begin{eqnarray}
& (TT)[i]^{s_1 N_1 I_1 ; s_2 N_2 I_2 ; j_M}_{j_L N_L I_L \alpha_L ; j_R N_R I_R \alpha_R} = \delta_{N_L+N_1+N_2,N_R} \delta_{I_L \otimes I_1 \otimes I_2, I_R} \nonumber \\
& \sqrt{2j_M+1} (-1)^{j_L+j_R+s_1+s_2} \sum\limits_j \left\{ \begin{array}{ccc} j_L & j_R & j \\ s_2 & s_1 & j_M \end{array} \right\} \nonumber \\
& \sqrt{2j+1} S[i]^{j(s_1 s_2) N_1 N_2 I_1 I_2}_{j_L N_L I_L \alpha_L ; j_R N_R I_R \alpha_R}. \label{StoTT}
\end{eqnarray}
Per group of $\left\{ j_M, N_M = N_L + N_1, I_M = I_L \otimes I_1 \right\}$, we can perform a singular value decomposition:
\begin{eqnarray}
& (TT)[i]^{s_1 N_1 I_1 ; s_2 N_2 I_2 ; j_M}_{j_L N_L I_L \alpha_L ; j_R N_R I_R \alpha_R} = \sum\limits_{\alpha_M} U[i]^{j_M N_M I_M}_{(j_L N_L I_L \alpha_L s_1 N_1 I_1);\alpha_M} \nonumber \\
& \lambda[i]^{j_M N_M I_M}_{\alpha_M} \left( \sqrt{\frac{2j_M+1}{2j_R+1}} V[i]^{j_M N_M I_M}_{\alpha_M;(j_R N_R I_R \alpha_R s_2 N_2 I_2)} \right).
\end{eqnarray}
After reshaping the indices to the normal form, it can be checked that $U[i]$ is the reduced part of a left-normalized MPS site tensor and that the term between brackets is the reduced part of a right-normalized MPS site tensor. The relation between $\lambda[i]$ and $\kappa[i]$ is given by
\begin{equation}
\kappa[i]_{j_M N_M I_M \alpha_M} = \frac{\lambda[i]_{j_M N_M I_M \alpha_M}}{\sqrt{ \sum\limits_{j_Q N_Q I_Q \alpha_Q} (2j_Q+1) \lambda[i]_{j_Q N_Q I_Q \alpha_Q}^2 }}.
\end{equation}
The $D_i$ largest values of $\lambda[i]$ are kept.

\section{CheMPS2 library} \label{ourcode}
CheMPS2 can be obtained from the CPC Program Library, and from its public git repository \cite{CheMPS2github}. The source code contains comments in Doxygen format. A complete reference manual can be generated from these comments. See \texttt{README} on how to install the library and on how to generate the manual. In this section, we give an overview of the basic structure of CheMPS2 so that new users can easily understand and alter the test runs to their own needs.

\subsection{The Hamiltonian}
Most molecular electronic structure programs have the ability to print matrix elements or to save them in binary format. CheMPS2 requires two-body matrix elements with eightfold permutation symmetry, which do not break $\mathsf{SU(2)}$ total electronic spin. A \texttt{CheMPS2::Hamiltonian} object should be created at the beginning of a calculation, and filled with the matrix elements of the problem at hand.

Users can utilize their preferred molecular electronic structure program to generate the matrix elements. The functions \texttt{setEconst}, \texttt{setTmat}, and \texttt{setVmat} then fill the \texttt{CheMPS2::Hamiltonian} object elementwise. Note that for $(ij | V | kl) = V_{ijkl}$ we have assumed the physics notation. This means that orbital $k$ at position $r_1$ (denoted by $k(r_1)$) scatters from orbital $l(r_2)$ into orbitals $i(r_1)$ and $j(r_2)$.

We have used Psi4 \cite{Psi4article} to generate molecular orbital matrix elements. Two plugins can be found in the folder \texttt{mointegrals}, with corresponding instructions in \texttt{README}. One plugin allows to print matrix elements as text during a Psi4 calculation, in a format which CheMPS2 is able to read. The other plugin creates a   \texttt{CheMPS2::Hamiltonian} object during a Psi4 calculation, fills it with the molecular orbital matrix elements, and stores it to disk in binary format. The latter option requires linking of the CheMPS2 library to the Psi4 plugin, but allows for reduced storage requirements.

In the \texttt{CheMPS2::Problem} object, users can specify the symmetry sector to which the calculations are restricted. The \texttt{CheMPS2::Hamiltonian} and the desired total electronic spin, particle number, and point group symmetry then completely determine a FCI calculation. In order to do DMRG or DMRG-SCF instead of resp. FCI or CASSCF, a convergence scheme for the subsequent sweeps should be set up.

\subsection{Convergence scheme} \label{ConvergenceSchemeSection}
The \texttt{CheMPS2::ConvergenceScheme} object controls the DMRG sweeps. It is divided into a number of consecutive instructions. Each instruction contains four parameters: the number of reduced renormalized basis states $D$ which should be kept, an energy threshold $E_{conv}$ for convergence, the maximum number of sweeps $N_{max}$, and the noise prefactor $\gamma_{noise}$.

The parameters $\gamma_{noise}$ and $D$ are relevant for the micro-iterations. Just before the decomposition of the reduced $S[i]$-tensor, random noise is added to it. This random noise is bounded in magnitude by $0.5 \gamma_{noise} w^{disc}(D)$, where $w^{disc}(D)$ is the maximum discarded weight obtained during the previous left- or right-sweep. After decomposition of the reduced $S[i]$-tensor, its reduced Schmidt spectrum $\lambda[i]$ is truncated to $D$.

The parameters $E_{conv}$ and $N_{max}$ are relevant for the macro-iterations. If after one macro-iteration (left- plus right-sweep), the energy difference is smaller than $E_{conv}$, the sweeping stops and the next instruction is performed. If energy convergence is not reached after $N_{max}$ macro-iterations, the current instruction ends as well.

\subsection{DMRG}
Creation of a \texttt{CheMPS2::DMRG} object requires a \texttt{CheMPS2::Hamiltonian}, a \texttt{CheMPS2::Problem}, and a \texttt{CheMPS2::ConvergenceScheme}. Each DMRG calculation starts by creating a new MPS. Its virtual dimension $D$ is obtained from the first instruction of the \texttt{CheMPS2::ConvergenceScheme} object. At each MPS bond, this virtual dimension $D$ is distributed equally over all possible symmetry sectors, ensuring that the dimension of a certain symmetry sector does not exceed the corresponding FCI dimension. The so-created MPS is filled with random noise.

The function \texttt{Solve} performs the instructions of the convergence scheme. Afterwards, it returns the minimal variational energy encountered during all the performed micro-iterations.

With the function \texttt{calc2DM}, the reduced 2-RDMs $\Gamma^A$ and $\Gamma^B$ are calculated:
\begin{eqnarray}
\Gamma_{(i \sigma) (j \tau) ; (k \sigma) (l \tau)} & = & \braket{ \hat{a}^{\dagger}_{i \sigma} \hat{a}^{\dagger}_{j \tau}  \hat{a}_{l \tau} \hat{a}_{k \sigma}}\\
\Gamma^{A}_{ij ; kl} & = & \sum\limits_{\sigma \tau} \Gamma_{(i \sigma) (j \tau) ; (k \sigma) (l \tau)} \\
\Gamma^{B}_{ij ; kl} & = & \sum\limits_{\sigma \tau} (-1)^{\sigma - \tau} \Gamma_{(i \sigma) (j \tau) ; (k \sigma) (l \tau)}
\end{eqnarray}
$\Gamma^A$ can be used to calculate the energy, the particle number $N$, and the 1-RDM:
\begin{eqnarray}
E & = & E_{const} + \frac{1}{2} \sum\limits_{ijkl} \Gamma^{A}_{ij ; kl} (ij | h | kl) \\
N(N-1) & = & \sum\limits_{ij} \Gamma^{A}_{ij ; ij} \\
\sum\limits_{\sigma} \braket{ \hat{a}^{\dagger}_{i \sigma} \hat{a}_{k \sigma}} & = & \frac{1}{N-1} \sum\limits_j \Gamma^{A}_{ij ; kj}
\end{eqnarray}
and is needed for the DMRG-SCF algorithm, while $\Gamma^B$ is important for spin-spin correlation functions.

The \texttt{CheMPS2::DMRG} object can also calculate excited states. After the ground state $\ket{\Psi_0}$ has been determined, the desired number of excited states can be set once with the function \texttt{activateExcitations}. Before \texttt{Solve} is called to find the next new excitation $\ket{\Psi_{m}}$, the function \texttt{newExcitation} should be called with the parameter $\eta_{m}$. This pushes back the current MPS which represents $\ket{\Psi_{m-1}}$, and sets the Hamiltonian to
\begin{equation}
 \hat{H}_m = \hat{H}_0 + \sum\limits_{k={0}}^{m-1} \eta_{k+1} \ket{\Psi_k} \bra{\Psi_k}.
\end{equation}
Our excited state DMRG algorithm is hence a state-specific algorithm, which projects out lower-lying states in the given $\mathsf{SU(2)} \otimes \mathsf{U(1)} \otimes \mathsf{P}$ symmetry sector. An example can be found in \texttt{tests/test5.cpp}.

OpenMP parallelization is used in the \texttt{CheMPS2::DMRG} object to speed up \textit{(a)} contractions involving tensors with a sparse block structure, for example the action of the effective Hamiltonian on a particular guess, and \textit{(b)} the construction of the (often similar) renormalized operators in between two micro-iterations.

\subsection{DMRG-SCF}
A state-specific DMRG-SCF algorithm is implemented in \texttt{CheMPS2::CASSCF}. Its creation requires a \texttt{CheMPS2::Hamiltonian} object. The number of occupied, active, and virtual orbitals per point group irrep should be given with the function \texttt{setupStart} before calling the SCF routine.

The CASSCF routine which is implemented is the augmented Hessian \cite{LengsfieldPaper} Newton-Raphson method from Ref. \cite{SiegbahnPaper}, with exact Hessian. It can be called with the function \texttt{doCASSCFnewtonraphson}, which requires the targeted symmetry sector, the convergence scheme, and the targeted root for the state-specific algorithm. When the gradient for orbital rotations reaches a predefined threshold, the routine returns the converged DMRG-SCF energy. An example can be found in \texttt{tests/test6.cpp}.

\section{Carbon dimer} \label{C2}
\subsection{Introduction}
Despite its simplicity at first sight, the carbon dimer provides a rich source of interesting physics. The bond between the two carbon atoms is of the charge-shift type \cite{bondingConundrums, ChargeShiftnaturechem}. Its strength tempts chemists to classify it as a quadruple bond \cite{PhysRev.56.778, C2wu, vonRaguSchleyer19936387, weinhold2005valency, C2natchem, ANIE201208206}, and recent research indicates how this fourth bond can be interpreted \cite{dunningC2}. The $1s$ core correlation is significant \cite{TheoChemC2, C2chinesen}. The low-lying bond dissociation curves are quasi-degenerate, and avoided crossings occur between states with the same spin and $D_{\infty h}$ point group symmetry \cite{BoggioPasqua2000159, abrams2004full, Varandas}. This happens for example between the $X^1\Sigma_g^+$ and $B'^1 \Sigma_g^+$ states, and between the $c^3\Sigma_u^+$ and $2^3\Sigma_u^+$ states. Fortunately, relativistic effects are small \cite{Kokkin, JiangC2}.

Accurate data for the low-lying states, preferably at the FCI level of theory for a given basis set, are useful to assess the accuracy of approximate molecular electronic structure methods. The $X^1\Sigma_g^+$, $B^1 \Delta_g$, and $B'^1\Sigma_g^+$ bond dissociation curves of Ref. \cite{abrams2004full} at the frozen core FCI/6-31G* level of theory are utilized to this end in several works \cite{sherillC2comppaper, useofAbrams2, useofAbrams3, useofAbrams4}.

The 12 lowest states of the carbon dimer are $X^1\Sigma_g^+$, $a^3\Pi_u$, $b^3\Sigma_g^-$, $A^1\Pi_u$, $c^3\Sigma_u^+$, $B^1 \Delta_g$, $B'^1\Sigma_g^+$, $d^3\Pi_g$, $C^1\Pi_g$, $1^1\Sigma_u^-$, $1^3\Delta_u$, and $2^3\Sigma_u^+$ \cite{BoggioPasqua2000159}. In section \ref{C2resultsSec}, we present the bond dissociation curves of these states at the DMRG(28o, 12e, D$_{ \mathsf{ SU(2) }}$=2500)/cc-pVDZ level of theory.

To estimate the contribution of $1s$ core correlation to the $X^1\Sigma_g^+$ bond dissociation curve, we compare energies at the DMRG(28o, 12e, D$_{\mathsf{SU(2)}}$=2500)/cc-pVDZ, DMRG-SCF(26o, 8e, D$_{\mathsf{SU(2)}}$=2500)/cc-pVDZ, DMRG(36o, 12e, D$_{\mathsf{SU(2)}}$=2500)/cc-pCVDZ, and DMRG-SCF(34o, 8e, D$_{\mathsf{SU(2)}}$=2500)/cc-pCVDZ levels of theory in section \ref{sec1sCoreCorr}. The cc-pCVDZ basis augments the cc-pVDZ basis with extra $1s$ and $1p$ functions to treat core and core-valence correlation \cite{ccpcvdzreference}. 

For all calculations, the variational energies are converged to $0.1 mE_h$ from the extrapolated value. This implies that, for all practical purposes, we present data at the FCI/cc-pVDZ, CASSCF(26o, 8e)/cc-pVDZ, FCI/cc-pCVDZ, and CASSCF(34o, 8e)/cc-pCVDZ levels of theory.

\subsection{Symmetry labeling}
\begin{figure}[t!]
 \includegraphics[width=0.45\textwidth]{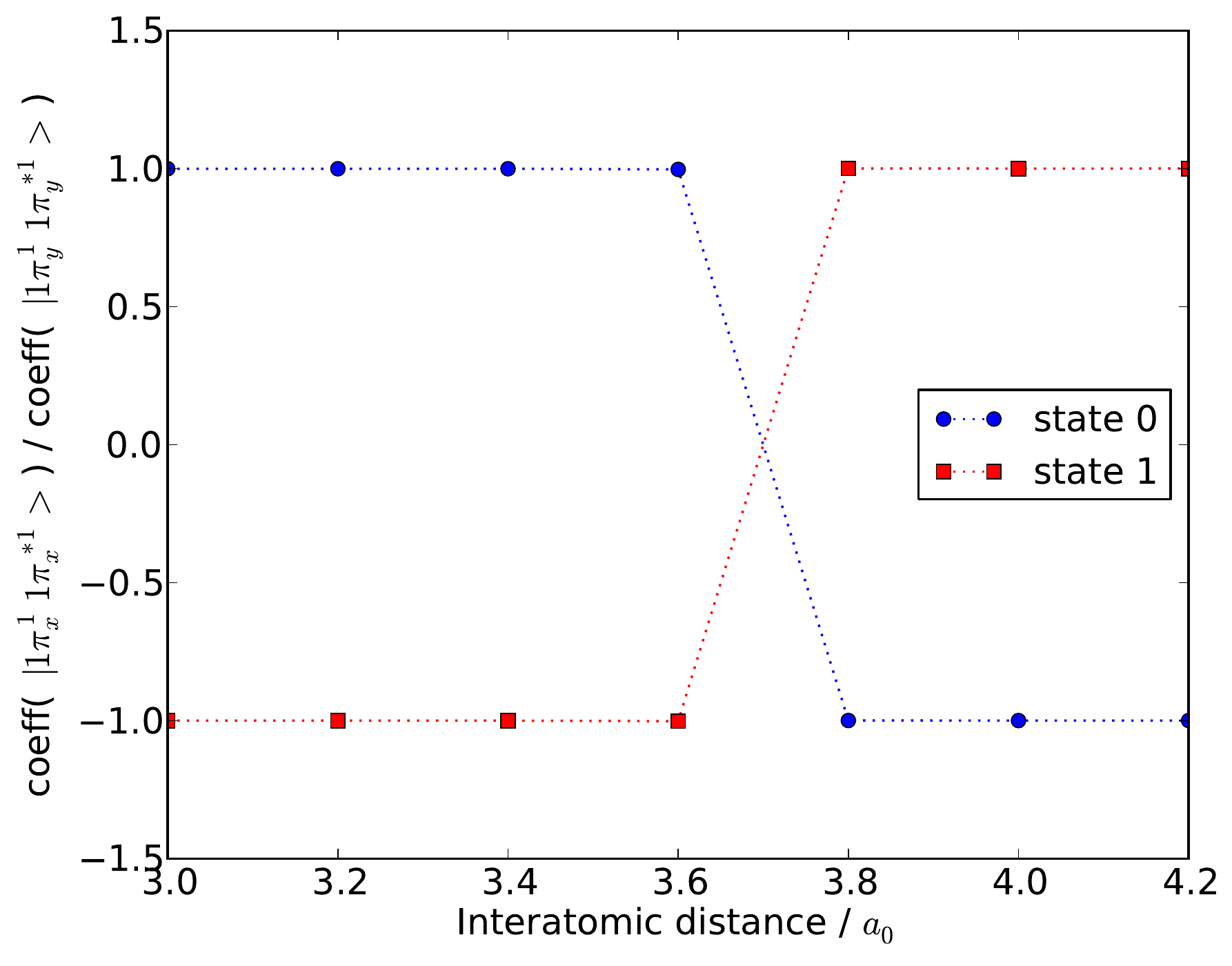}
 \caption{\label{fig3B1ucoeff} For the cc-pVDZ basis, the $1^3\Delta_u$ state drops below the $c^3\Sigma_u^+$ state at an interatomic distance between 3.6 and 3.8 $a_0$. The $\ket{1\pi_x^1 1\pi_x^{*1}}$ and $\ket{1\pi_y^1 1\pi_y^{*1}}$ FCI coefficients allow to correctly label the $^3B_{1u}$ ground state (state 0) and first excited state (state 1).}
\end{figure}

Since CheMPS2 can only handle abelian point groups, we use $D_{2h}$ point group symmetry to obtain these 12 states:
\begin{eqnarray}
X^1\Sigma_g^+ ; B^1 \Delta_g ; B'^1\Sigma_g^+ & \rightarrow & ^1A_g \\
c^3\Sigma_u^+ ; 1^3\Delta_u ; 2^3\Sigma_u^+ &\rightarrow & ^3B_{1u} \\
C^1\Pi_g & \rightarrow & ^1B_{2g} \label{firstOfThed2hIrreps}\\
A^1\Pi_u & \rightarrow & ^1B_{2u} \\
1^1\Sigma_u^- & \rightarrow & ^1A_u \\
b^3 \Sigma_g^- & \rightarrow & ^3B_{1g} \\
d^3\Pi_g & \rightarrow & ^3B_{2g} \\
a^3\Pi_u &\rightarrow & ^3B_{2u}. \label{lastOfThed2hIrreps}
\end{eqnarray}
For the states (\ref{firstOfThed2hIrreps})-(\ref{lastOfThed2hIrreps}), we have calculated one extra state to check that no unexpected curve crossings occur. To discern the lowest three $^1A_g$ states, we have extracted the following FCI coefficients from the DMRG object \cite{abrams2004full}:
\begin{eqnarray}
\ket{1\pi_x^2} & = & \ket{1\sigma_g^2 1\sigma_u^2 2\sigma_g^2 2\sigma_u^2 \mathbf{1\pi_x^2} 3\sigma_g^2} \\
               & = & \ket{1A_g^2 1B_{1u}^2 2A_g^2 2B_{1u}^2 \mathbf{1B_{3u}^2} 3A_g^2} \\
\ket{1\pi_y^2} & = & \ket{1\sigma_g^2 1\sigma_u^2 2\sigma_g^2 2\sigma_u^2 \mathbf{1\pi_y^2} 3\sigma_g^2} \\
               & = & \ket{1A_g^2 1B_{1u}^2 2A_g^2 2B_{1u}^2 \mathbf{1B_{2u}^2} 3A_g^2}
\end{eqnarray}
When the FCI coefficients are equal, the state has $^1\Sigma_g^+$ symmetry, and when the FCI coefficients are each other's additive inverse, the state has $^1\Delta_g$ symmetry. To discern the lowest three $^3B_{1u}$ states, we have extracted the following FCI coefficients from the DMRG object:
\begin{eqnarray}
\ket{1\pi_x^1 1\pi_x^{*1}} & = & \ket{1\sigma_g^2 1\sigma_u^2 2\sigma_g^2 2\sigma_u^2 \mathbf{1\pi_x^1} 3\sigma_g^2 \mathbf{1\pi_x^{*1}}} \\
               & = & \ket{1A_g^2 1B_{1u}^2 2A_g^2 2B_{1u}^2 \mathbf{1B_{3u}^1} 3A_g^2 \mathbf{1B_{2g}^1}} \\
\ket{1\pi_y^1 1\pi_y^{*1}} & = & \ket{1\sigma_g^2 1\sigma_u^2 2\sigma_g^2 2\sigma_u^2 \mathbf{1\pi_y^1} 3\sigma_g^2 \mathbf{1\pi_y^{*1}}} \\
               & = & \ket{1A_g^2 1B_{1u}^2 2A_g^2 2B_{1u}^2 \mathbf{1B_{2u}^1} 3A_g^2 \mathbf{1B_{3g}^1}}
\end{eqnarray}
When the FCI coefficients are equal, the state has $^3\Sigma_u^+$ symmetry, and when the FCI coefficients are each other's additive inverse, the state has $^3\Delta_u$ symmetry. An example is shown in Fig. \ref{fig3B1ucoeff}.

\subsection{Irrep ordering} \label{secIrrepOrder}
\begin{figure}[t!]
 \includegraphics[width=0.45\textwidth]{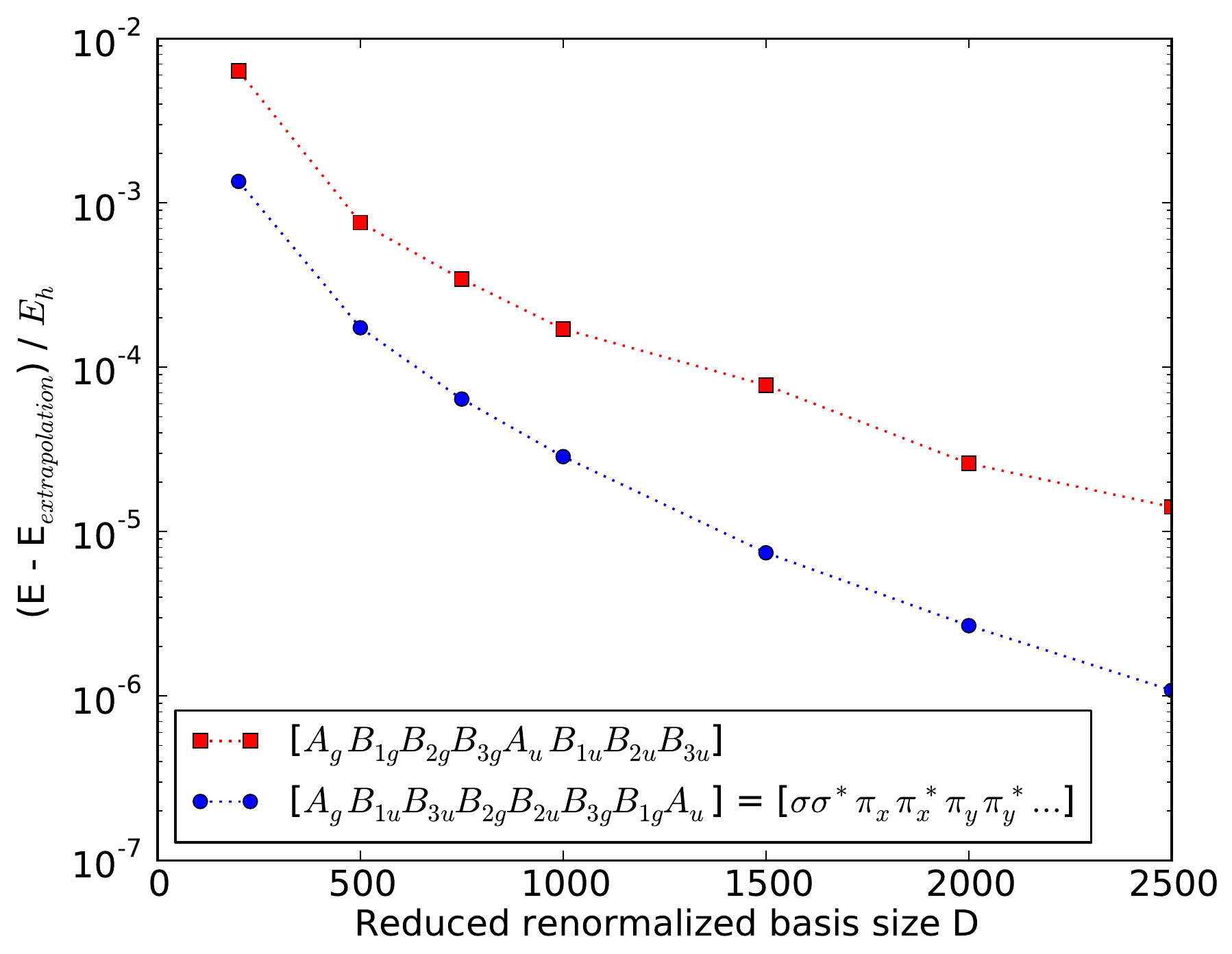}
 \caption{\label{figIrrepOrder} The orbital choice and ordering influence the convergence behaviour of DMRG. The convergence behaviour of two irrep orderings is shown for the carbon dimer with interatomic distance 2.4 $a_0$ in the cc-pVDZ basis. The extrapolated energy was obtained from the ordering where bonding and anti-bonding orbitals are grouped, with the method described in section \ref{HowToExtrapolateEnergies}.}
\end{figure}

The standard $D_{2h}$ irrep order is not optimal to study the carbon dimer with DMRG. As stated in section \ref{ConvergenceSection}, it is best to group bonding and anti-bonding orbitals together on the DMRG lattice. The convergence behaviour of these two irrep orderings is shown in Fig. \ref{figIrrepOrder}. We have used the latter ordering for our calculations.

\subsection{Extrapolation} \label{HowToExtrapolateEnergies}
\begin{figure}[t!]
 \includegraphics[width=0.45\textwidth]{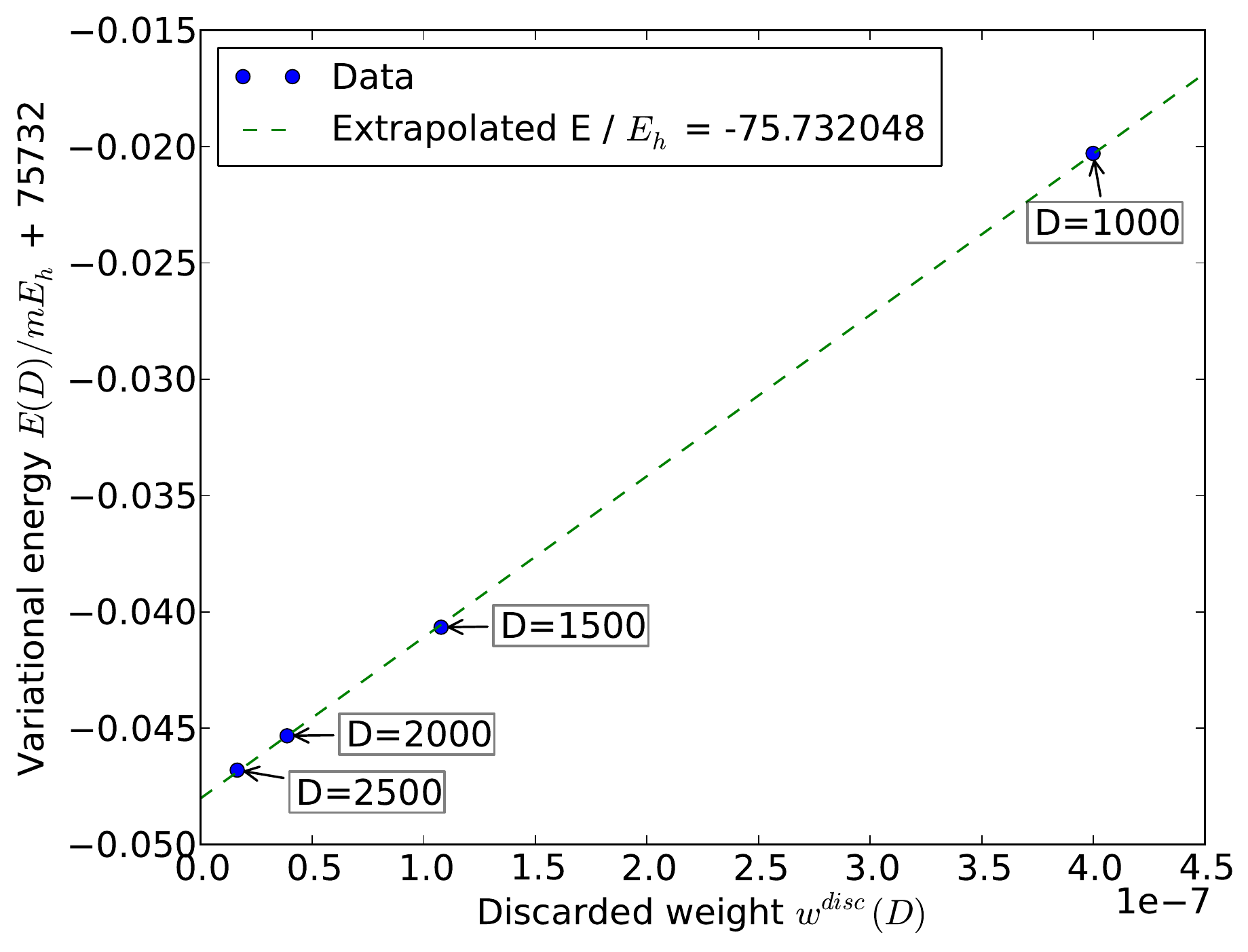}
 \caption{\label{figExtrapolationExample} The extrapolation scheme of Eq. (\ref{extrapolSchemeEq}) is used to obtain energies which are correct up to 0.01 $mE_h$. The example shown here is for the $X^1\Sigma_g^+$ state of the carbon dimer at an interatomic distance of 2.35 $a_0$ in the cc-pVDZ basis.}
\end{figure}

We have used the convergence scheme in Tab. \ref{tableC2convscheme} for all the calculations of the carbon dimer. The extrapolation scheme of Eq. (\ref{extrapolSchemeEq}) is used to obtain energies which are correct up to 0.01 $mE_h$. An example of such an extrapolation is shown in Fig. \ref{figExtrapolationExample}. The energies shown in sections \ref{C2resultsSec} and \ref{sec1sCoreCorr} are the extrapolated values.

\begin{table}
\caption{\label{tableC2convscheme} Convergence scheme for the carbon dimer calculations. The symbols are explained in section \ref{ConvergenceSchemeSection}.}
\begin{center}
\begin{tabular}{|cccc|}
\hline
$D_{\mathsf{SU(2)}}$ & $\gamma_{noise}$ & $E_{conv} / E_h$ & $N_{max}$\\
\hline
200 & 0.03 & $10^{-8}$ & 2 \\
200 & 0.00 & $10^{-8}$ & 3 \\
500 & 0.03 & $10^{-8}$ & 2 \\
500 & 0.00 & $10^{-8}$ & 5 \\
1000 & 0.03 & $10^{-8}$ & 2 \\
1000 & 0.00 & $10^{-8}$ & 5 \\
1500 & 0.03 & $10^{-8}$ & 2 \\
1500 & 0.00 & $10^{-8}$ & 5 \\
2000 & 0.03 & $10^{-8}$ & 2 \\
2000 & 0.00 & $10^{-8}$ & 5 \\
2500 & 0.03 & $10^{-8}$ & 2 \\
2500 & 0.00 & $10^{-8}$ & 12 \\
\hline
\end{tabular}
\end{center}
\end{table}

\begin{landscape}
\begin{table}
\caption{\label{largeTableC2} Extrapolated energies for the 12 lowest states of the carbon dimer at the DMRG(28o, 12e, D$_{\mathsf{SU(2)}}$=2500)/cc-pVDZ level of theory. The energies are shifted 75 $E_h$ upwards, and are expressed in $mE_h$.}
\begin{center}
\begin{tabular}{|l|rrrrrrrrrrrr|}
\hline
 & \multicolumn{12}{|c|}{(Energy + 75 $E_h$) / $mE_h$}\\
R / $a_0$ & $X^1\Sigma_g^+$ & $a^3\Pi_u$ & $b^3\Sigma_g^-$ & $A^1\Pi_u$ & $c^3\Sigma_u^+$ & $B^1 \Delta_g$ & $B'^1\Sigma_g^+$ & $d^3\Pi_g$ & $C^1\Pi_g$ &  $1^1\Sigma_u^-$ & $1^3\Delta_u$ & $2^3\Sigma_u^+$ \\
\hline
1.8 &  -454.96  &  -357.88  &  -253.42  &  -314.72  &  -439.01  &  -207.85  &  -263.42  &  -311.47  &  -250.35  &  -4.51  &  -35.05  &  -70.74  \\
1.9 &  -562.08  &  -485.42  &  -396.48  &  -442.69  &  -541.14  &  -353.53  &  -381.99  &  -430.96  &  -368.18  &  -145.91  &  -177.61  &  -212.67  \\
2.0 &  -635.85  &  -576.98  &  -501.10  &  -534.76  &  -609.66  &  -460.60  &  -471.77  &  -514.53  &  -449.98  &  -251.39  &  -284.10  &  -318.58  \\
2.1 &  -684.30  &  -640.94  &  -576.18  &  -599.30  &  -652.70  &  -537.96  &  -538.97  &  -570.80  &  -504.41  &  -329.17  &  -362.69  &  -396.52  \\
2.2 &  -713.63  &  -683.80  &  -628.60  &  -642.81  &  -676.60  &  -592.52  &  -587.67  &  -606.46  &  -538.20  &  -385.65  &  -419.78  &  -452.75  \\
2.3 &  -728.68  &  -710.64  &  -663.75  &  -670.34  &  -686.31  &  -629.65  &  -621.40  &  -626.70  &  -556.61  &  -425.92  &  -460.33  &  -492.17  \\
2.35 &  -732.05  &  -719.33  &  -676.19  &  -679.39  &  -687.10  &  -643.04  &  -633.65  &  -632.36  &  -561.40  &  -441.36  &  -475.66  &  -506.80  \\
2.4 &  -733.18  &  -725.42  &  -685.81  &  -685.86  &  -685.73  &  -653.57  &  -643.30  &  -635.62  &  -563.85  &  -454.45  &  -488.25  &  -518.58  \\
2.5 &  -730.05  &  -731.22  &  -698.04  &  -692.42  &  -677.93  &  -667.55  &  -656.08  &  -636.43  &  -563.36  &  -477.13  &  -506.68  &  -534.93  \\
2.6 &  -721.58  &  -730.43  &  -702.98  &  -692.43  &  -665.39  &  -674.15  &  -661.94  &  -631.72  &  -558.20  &  -499.94  &  -518.21  &  -543.51  \\
2.7 &  -709.54  &  -724.91  &  -702.58  &  -687.72  &  -650.09  &  -675.32  &  -662.63  &  -623.62  &  -551.37  &  -519.32  &  -525.47  &  -546.10  \\
2.8 &  -695.37  &  -716.10  &  -698.35  &  -679.74  &  -633.70  &  -672.60  &  -659.48  &  -613.89  &  -545.69  &  -533.86  &  -532.15  &  -544.31  \\
2.9 &  -680.23  &  -705.08  &  -691.43  &  -669.58  &  -617.56  &  -667.13  &  -653.44  &  -603.99  &  -542.36  &  -544.27  &  -539.84  &  -541.16  \\
3.0 &  -665.20  &  -692.69  &  -682.70  &  -658.08  &  -602.65  &  -659.80  &  -645.08  &  -594.90  &  -540.48  &  -551.39  &  -546.21  &  -543.11  \\
3.2 &  -638.95  &  -666.17  &  -662.28  &  -633.46  &  -578.29  &  -642.09  &  -622.59  &  -579.90  &  -537.09  &  -558.59  &  -553.13  &  -549.22  \\
3.4 &  -617.95  &  -639.87  &  -640.64  &  -609.35  &  -561.37  &  -623.07  &  -597.29  &  -567.38  &  -532.80  &  -559.79  &  -554.31  &  -549.69  \\
3.6 &  -599.65  &  -615.55  &  -619.67  &  -587.68  &  -552.43  &  -604.72  &  -575.15  &  -556.01  &  -528.47  &  -557.69  &  -552.17  &  -544.05  \\
3.8 &  -583.60  &  -594.03  &  -600.33  &  -569.45  &  -547.56  &  -588.06  &  -557.98  &  -546.10  &  -525.09  &  -553.95  &  -548.30  &  -536.60  \\
4.0 &  -569.91  &  -575.68  &  -583.08  &  -555.06  &  -542.97  &  -573.57  &  -545.70  &  -538.27  &  -523.05  &  -549.57  &  -543.76  &  -531.21  \\
4.2 &  -558.63  &  -560.66  &  -568.16  &  -544.44  &  -538.59  &  -561.46  &  -537.47  &  -532.75  &  -522.22  &  -545.17  &  -539.23  &  -527.76  \\
4.4 &  -549.67  &  -548.99  &  -555.69  &  -537.12  &  -534.71  &  -551.75  &  -532.20  &  -529.22  &  -522.24  &  -541.13  &  -535.16  &  -525.67  \\
4.6 &  -542.81  &  -540.54  &  -545.74  &  -532.36  &  -531.58  &  -544.27  &  -528.91  &  -527.13  &  -522.69  &  -537.63  &  -531.81  &  -524.43  \\
4.8 &  -537.73  &  -534.90  &  -538.26  &  -529.39  &  -529.24  &  -538.70  &  -526.87  &  -525.96  &  -523.30  &  -534.74  &  -529.26  &  -523.70  \\
5.0 &  -534.05  &  -531.40  &  -533.02  &  -527.59  &  -527.64  &  -534.66  &  -525.60  &  -525.34  &  -523.89  &  -532.41  &  -527.46  &  -523.27  \\
5.2 &  -531.41  &  -529.29  &  -529.61  &  -526.50  &  -526.56  &  -531.78  &  -524.80  &  -525.01  &  -524.38  &  -530.57  &  -526.25  &  -523.03  \\
5.4 &  -529.51  &  -528.01  &  -527.51  &  -525.82  &  -525.87  &  -529.72  &  -524.29  &  -524.84  &  -524.73  &  -529.13  &  -525.48  &  -522.93  \\
5.6 &  -528.14  &  -527.19  &  -526.27  &  -525.38  &  -525.42  &  -528.23  &  -523.96  &  -524.73  &  -524.96  &  -528.00  &  -524.99  &  -522.90  \\
5.8 &  -527.13  &  -526.62  &  -525.53  &  -525.08  &  -525.10  &  -527.15  &  -523.75  &  -524.65  &  -525.08  &  -527.12  &  -524.68  &  -522.93  \\
6.0 &  -526.36  &  -526.20  &  -525.08  &  -524.87  &  -524.87  &  -526.38  &  -523.61  &  -524.58  &  -525.12  &  -526.43  &  -524.49  &  -522.99  \\
\hline

\end{tabular}
\end{center}
\end{table}
\end{landscape}

\subsection{Bond dissociation curves} \label{C2resultsSec}
\begin{figure}[t!]
 \includegraphics[width=0.45\textwidth]{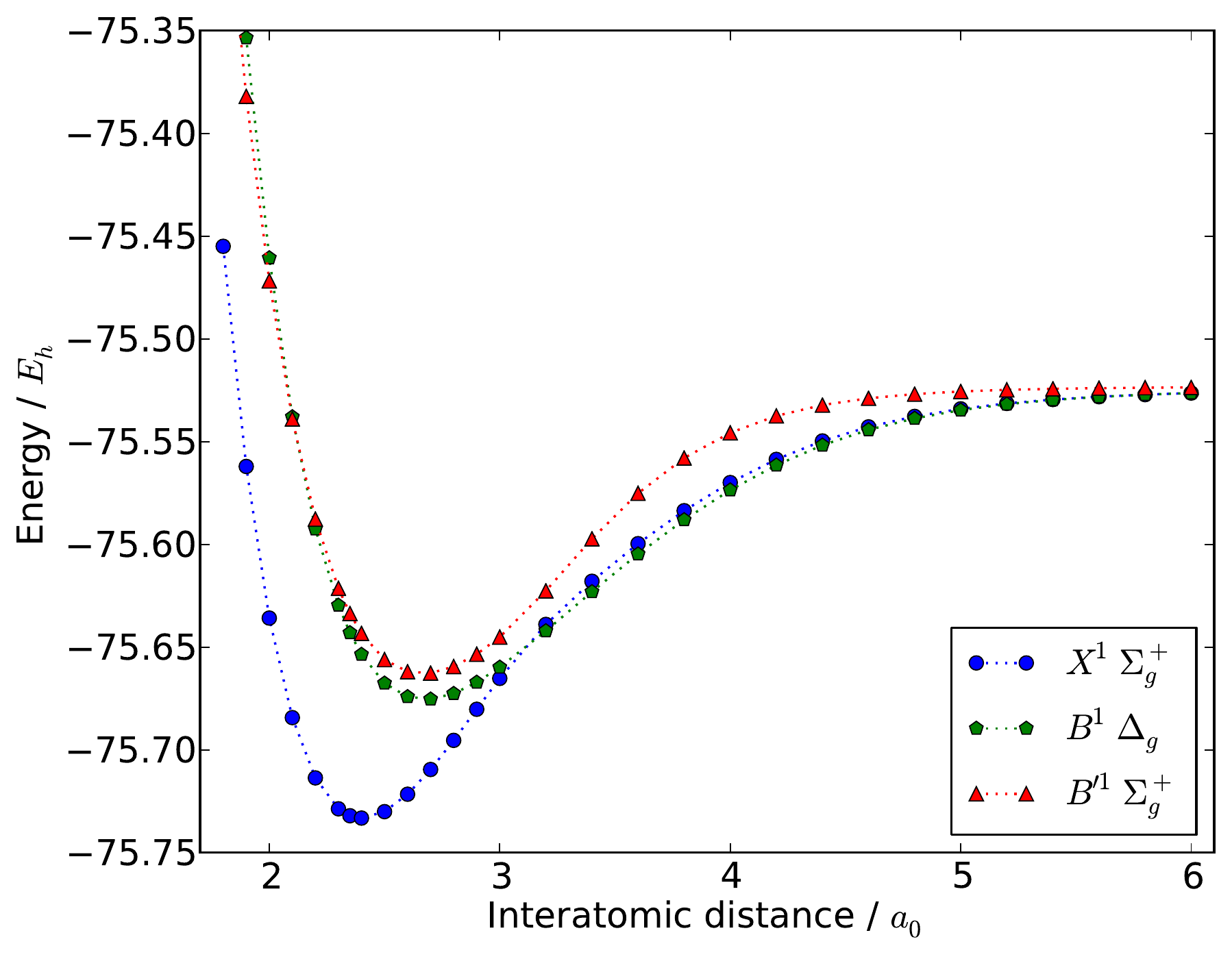}
 \caption{\label{1AgStates} Bond dissociation curves for the low-lying $^1A_g$ states of the carbon dimer in the cc-pVDZ basis.}
\end{figure}
\begin{figure}[t!]
 \includegraphics[width=0.45\textwidth]{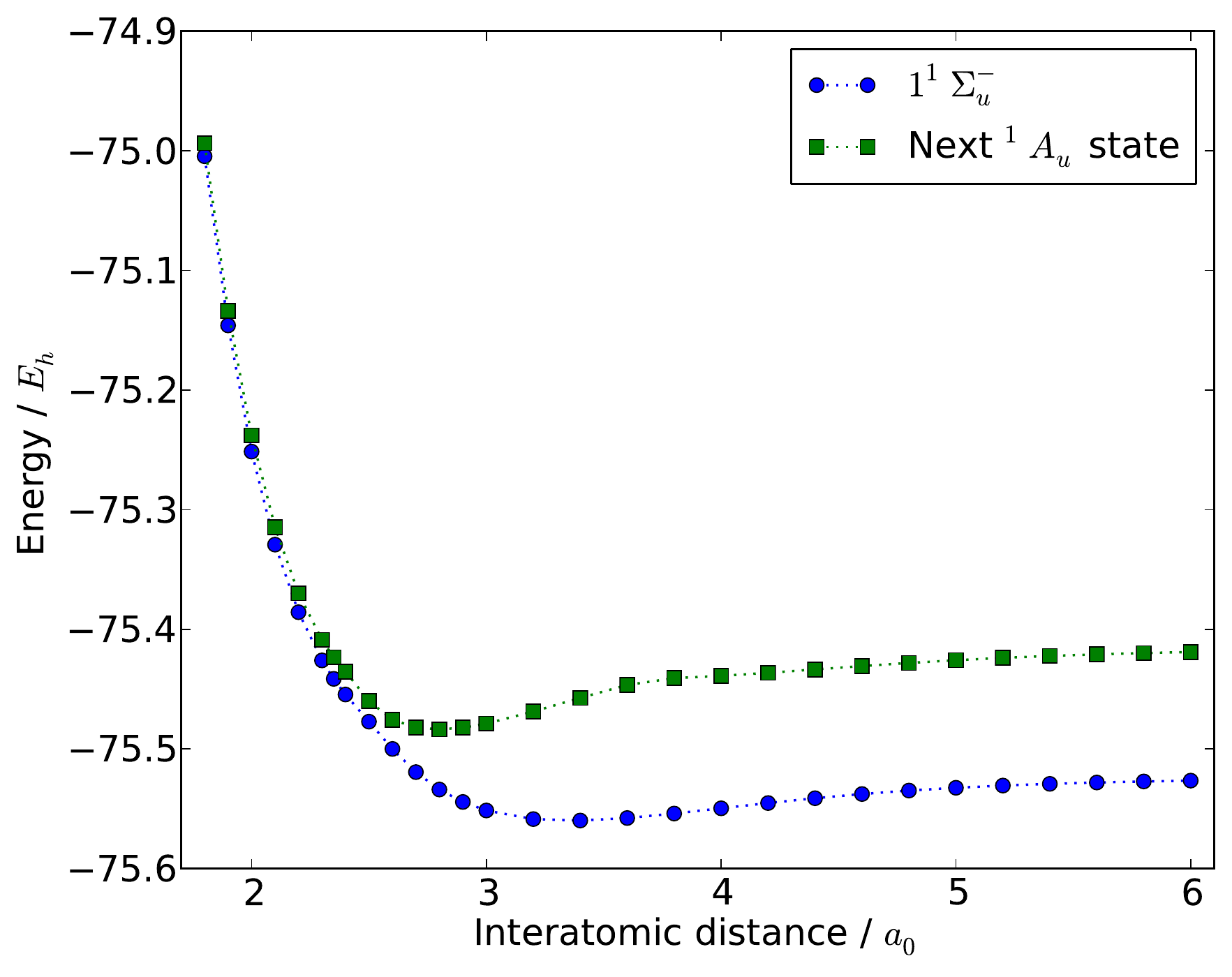}
 \caption{\label{1AuStates} Bond dissociation curves for the low-lying $^1A_u$ states of the carbon dimer in the cc-pVDZ basis.}
\end{figure}
\begin{figure}[t!]
 \includegraphics[width=0.45\textwidth]{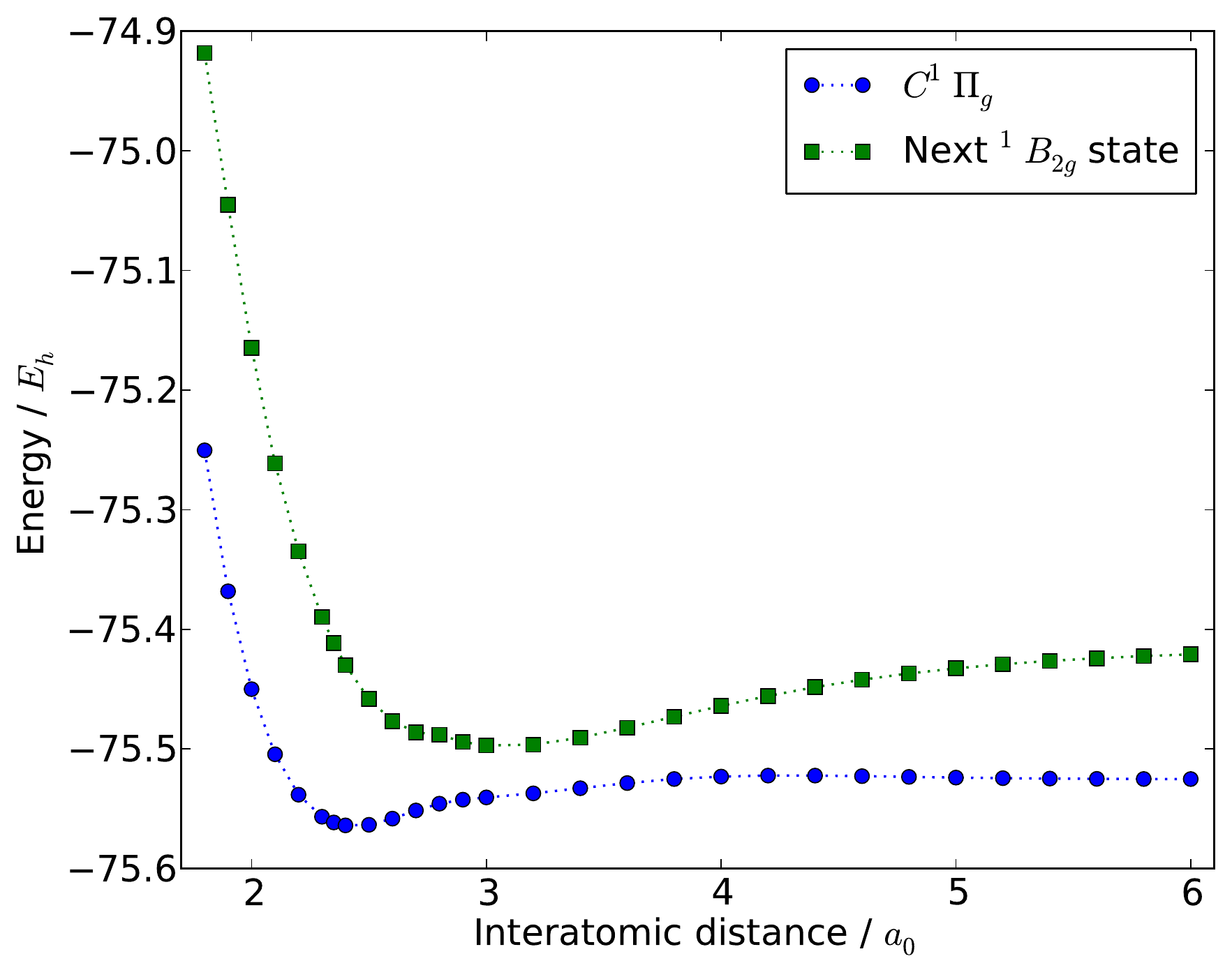}
 \caption{\label{1B2gStates} Bond dissociation curves for the low-lying $^1B_{2g}$ states of the carbon dimer in the cc-pVDZ basis.}
\end{figure}
\begin{figure}[t!]
 \includegraphics[width=0.45\textwidth]{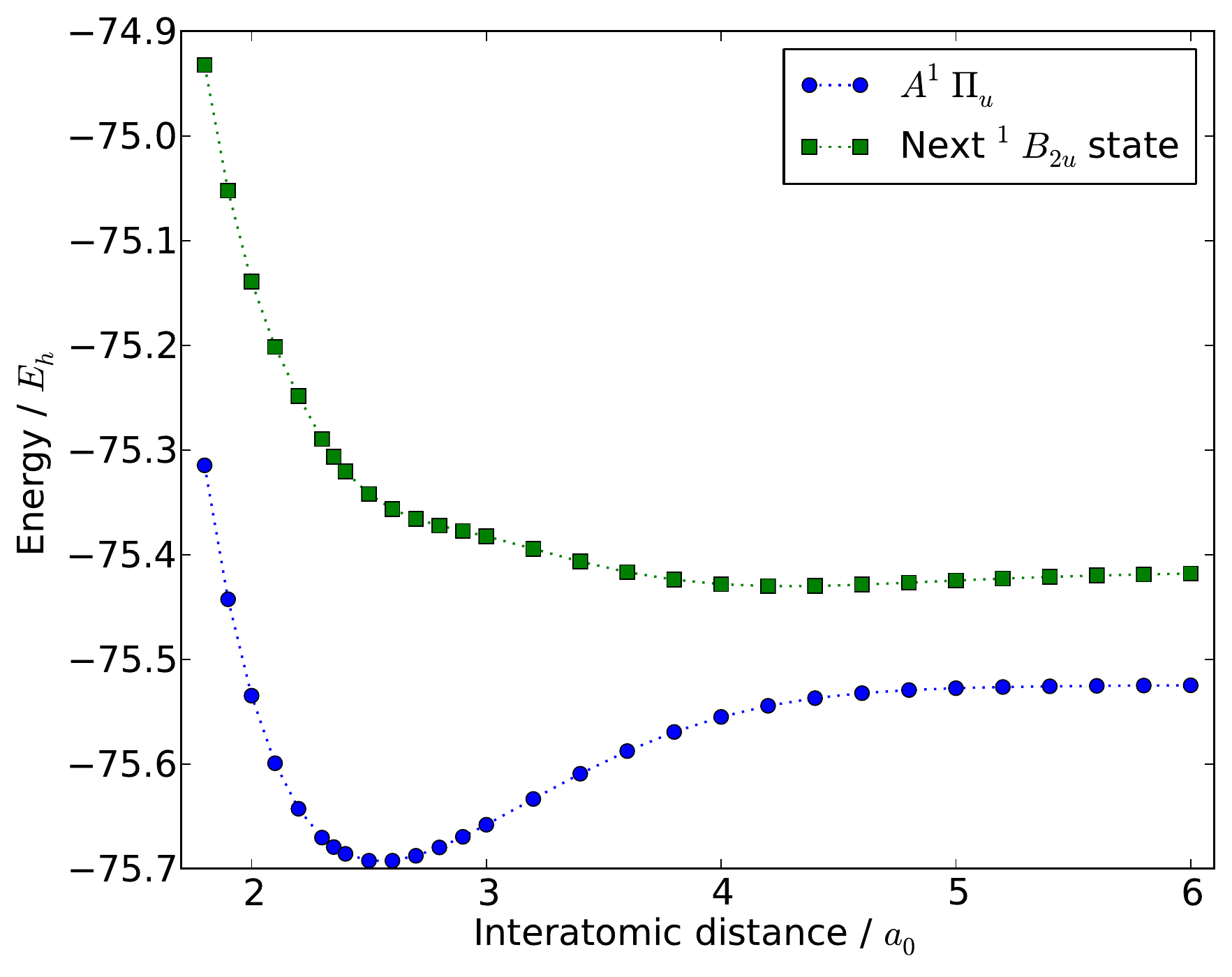}
 \caption{\label{1B2uStates} Bond dissociation curves for the low-lying $^1B_{2u}$ states of the carbon dimer in the cc-pVDZ basis.}
\end{figure}
\begin{figure}[t!]
 \includegraphics[width=0.45\textwidth]{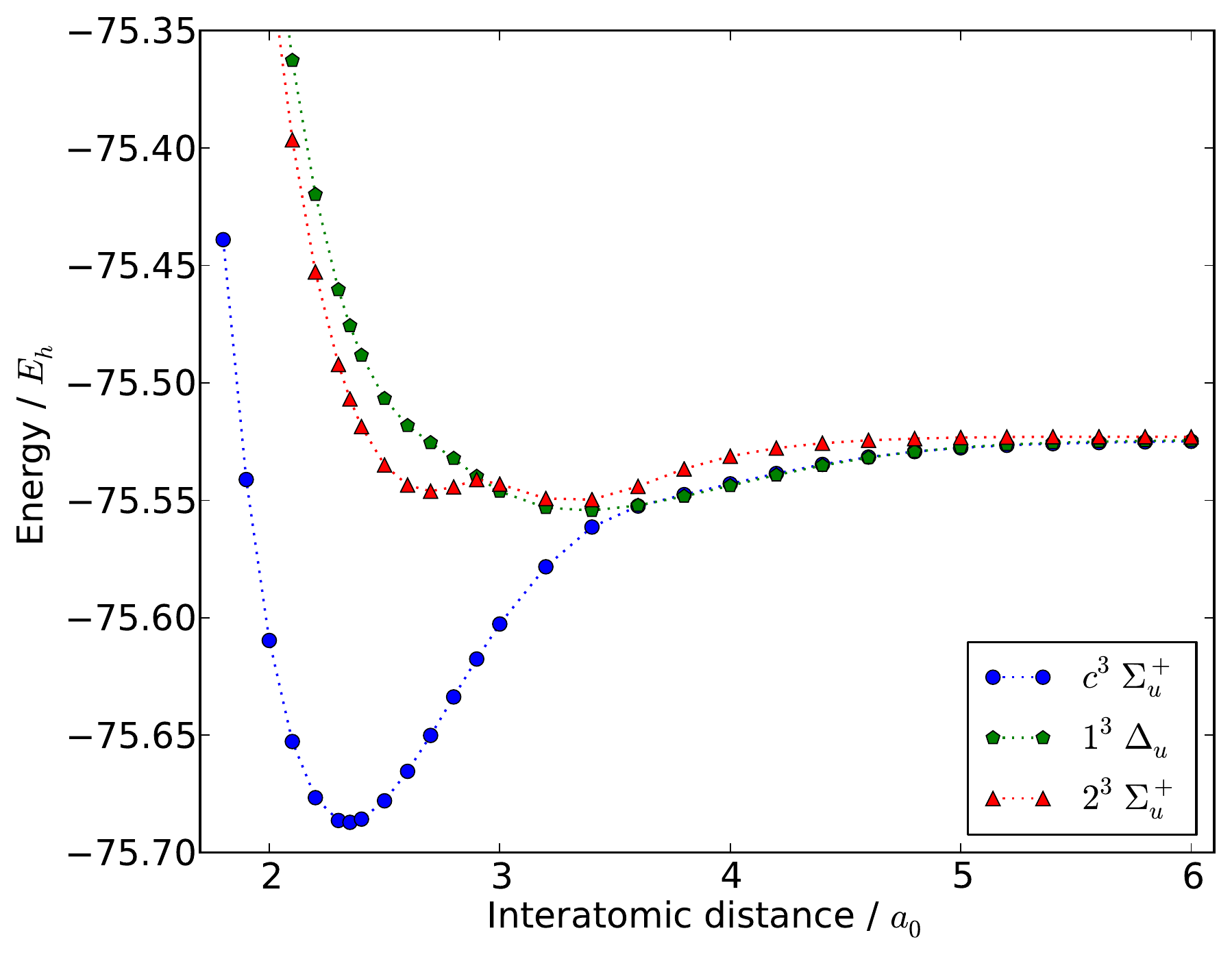}
 \caption{\label{3B1uStates} Bond dissociation curves for the low-lying $^3B_{1u}$ states of the carbon dimer in the cc-pVDZ basis.}
\end{figure}
\begin{figure}[t!]
 \includegraphics[width=0.45\textwidth]{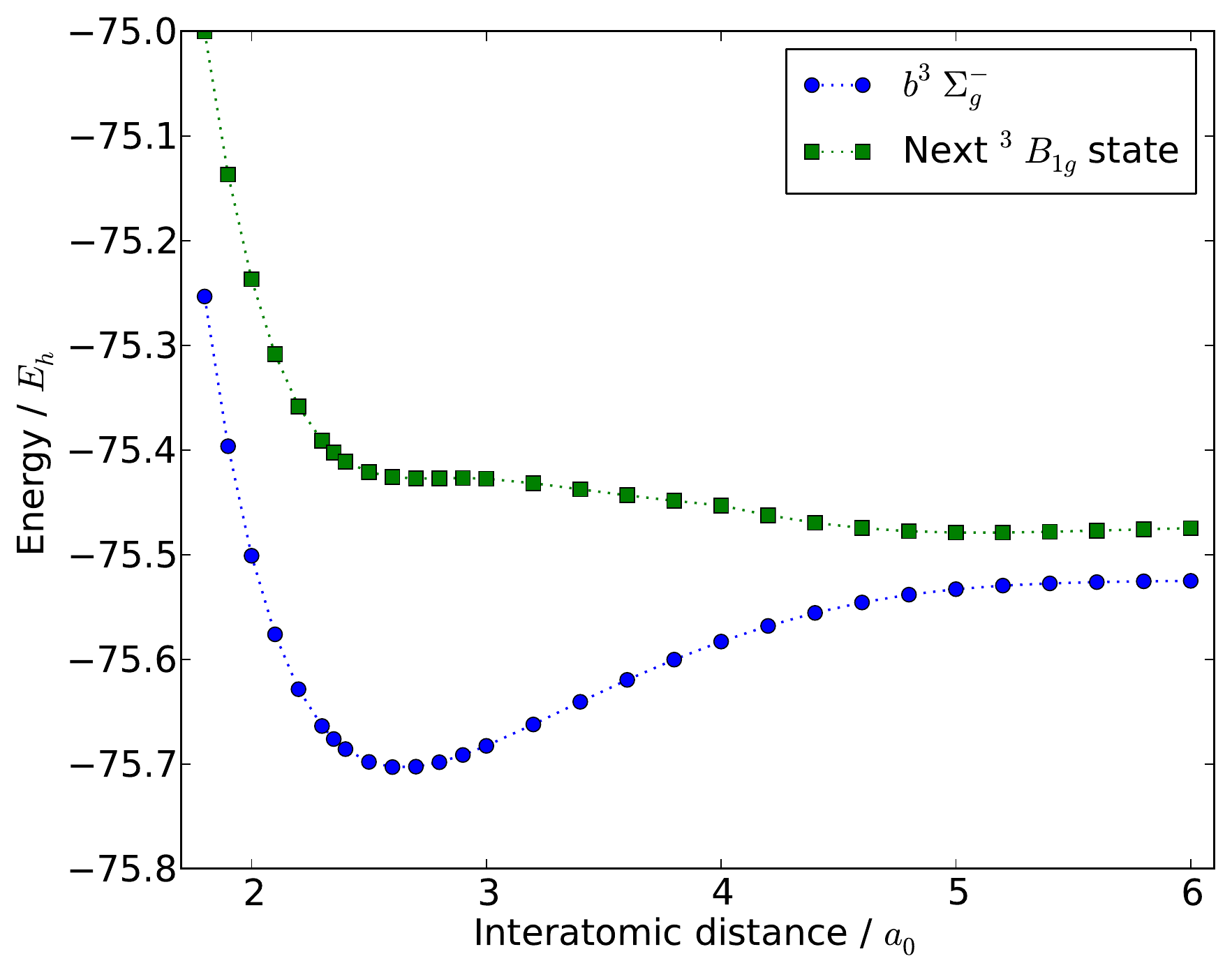}
 \caption{\label{3B1gStates} Bond dissociation curves for the low-lying $^3B_{1g}$ states of the carbon dimer in the cc-pVDZ basis.}
\end{figure}
\begin{figure}[t!]
 \includegraphics[width=0.45\textwidth]{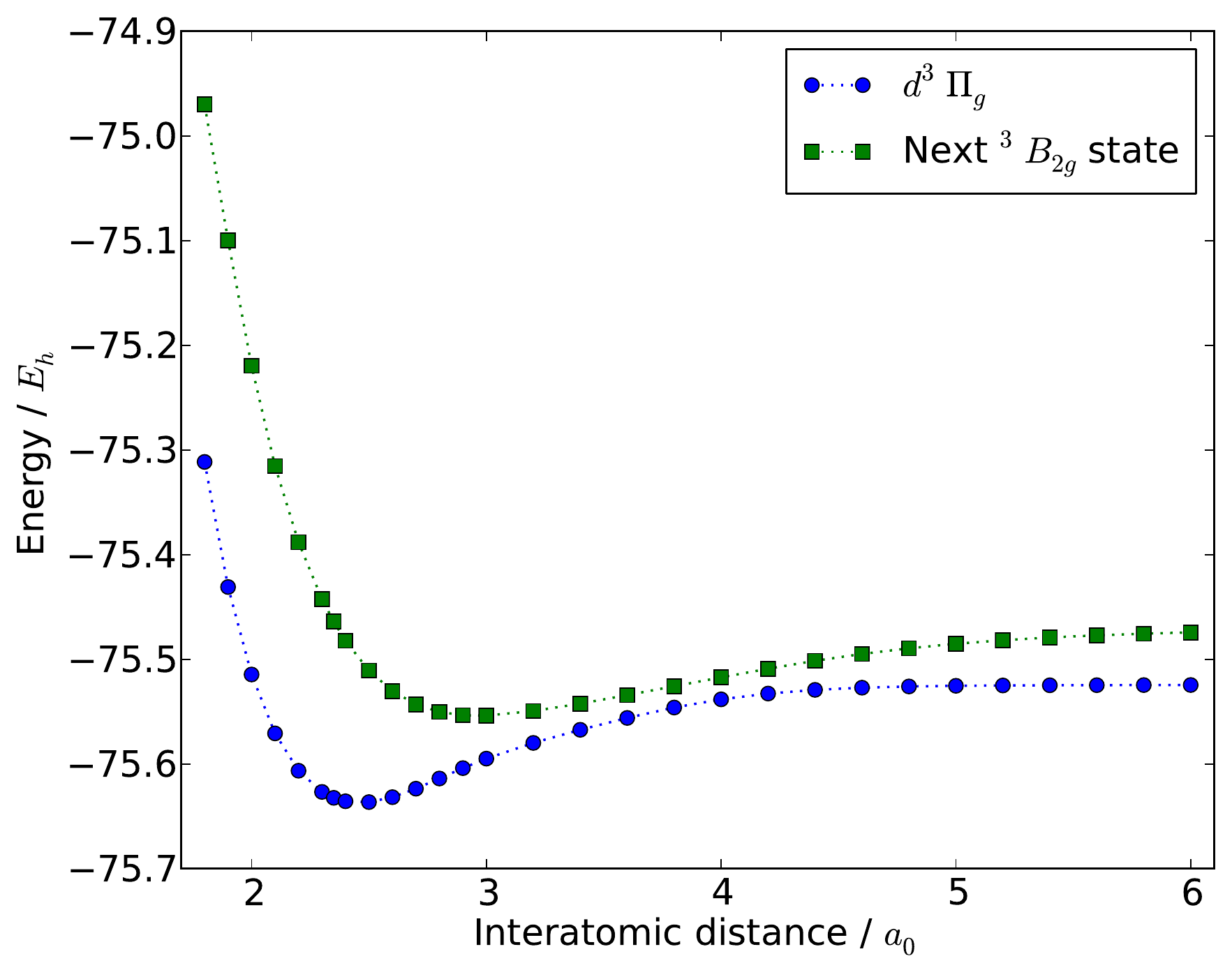}
 \caption{\label{3B2gStates} Bond dissociation curves for the low-lying $^3B_{2g}$ states of the carbon dimer in the cc-pVDZ basis.}
\end{figure}
\begin{figure}[t!]
 \includegraphics[width=0.45\textwidth]{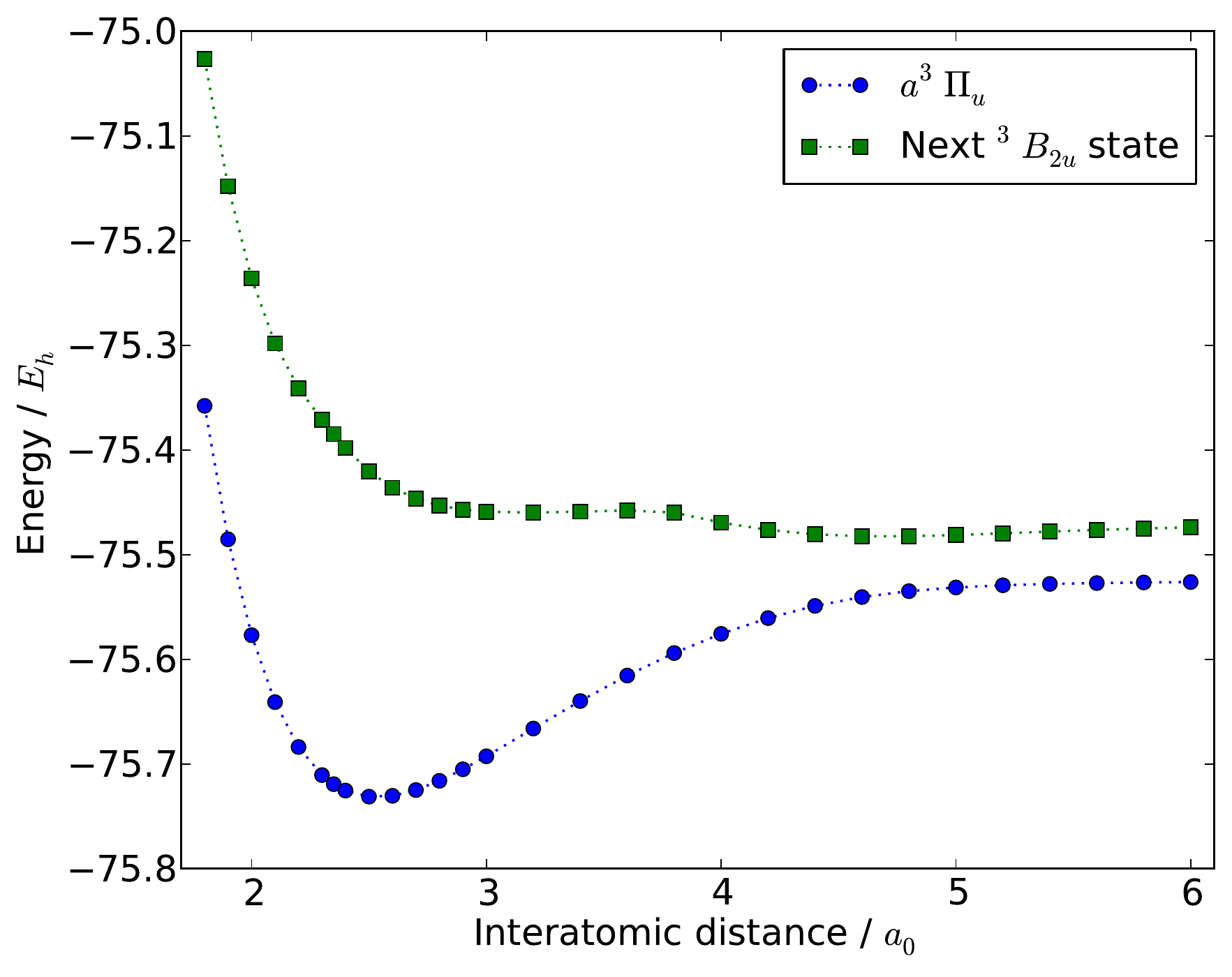}
 \caption{\label{3B2uStates} Bond dissociation curves for the low-lying $^3B_{2u}$ states of the carbon dimer in the cc-pVDZ basis.}
\end{figure}

The extrapolated energies at the DMRG(28o, 12e, D$_{\mathsf{SU(2)}}$=2500)/cc-pVDZ level of theory are summarized in Tab. \ref{largeTableC2} and are shown per targeted symmetry sector in Figs. \ref{1AgStates} to \ref{3B2uStates}. For the $^1A_g$ symmetry, the $B^1\Delta_g$ state drops below the $B'^1\Sigma_g^+$ state at an interatomic distance between $2 a_0$ and $2.1 a_0$, and it drops below the $X^1\Sigma_g^+$ state at an interatomic distance between $3a_0$ and $3.2 a_0$. The $X^1\Sigma_g^+$ and $B'^1\Sigma_g^+$ states have an avoided crossing. For the $^3B_{1u}$ symmetry, the $1^3\Delta_u$ state drops below the $2^3\Sigma_u^+$ state at an interatomic distance between $2.9 a_0$ and $3.0 a_0$, and it drops below the $c^3\Sigma_u^+$ state at an interatomic distance between $3.6 a_0$ and $3.8 a_0$. The $c^3\Sigma_u^+$ and $2^3\Sigma_u^+$ states have an avoided crossing. The intermediary peak of the $2^3\Sigma_u^+$ state near $2.9 a_0$ was also observed in Ref. \cite{BoggioPasqua2000159}, and is due to an avoided crossing with the $3^3\Sigma_u^+$ state. The $C^1\Pi_g$ and $d^3\Pi_g$ states also clearly show an avoided crossing with the next corresponding excited state.

\subsection{Core correlation} \label{sec1sCoreCorr}

\begin{table}
\caption{\label{CoreCorrTable} Extrapolated energies for the $X^1\Sigma_g^+$ state of the carbon dimer. (26o, 8e), (28o, 12e), (34o, 8e), and (36o, 12e) are shorthands for resp. DMRG-SCF(26o, 8e, D$_{\mathsf{SU(2)}}$=2500)/cc-pVDZ, DMRG(28o, 12e, D$_{\mathsf{SU(2)}}$=2500)/cc-pVDZ, DMRG-SCF(34o, 8e, D$_{\mathsf{SU(2)}}$=2500)/cc-pCVDZ, and DMRG(36o, 12e, D$_{\mathsf{SU(2)}}$=2500)/cc-pCVDZ. The energies are shifted 75 $E_h$ upwards, and are expressed in $mE_h$.}
\begin{center}
\begin{tabular}{|l|rrrr|}
\hline
 & \multicolumn{4}{|c|}{(Energy + 75 $E_h$) / $mE_h$}\\
R / $a_0$ & (26o, 8e) & (28o, 12e) & (34o, 8e) & (36o, 12e) \\
\hline
1.8  &  -450.44  &  -454.96  &  -459.72  &  -534.24  \\
1.9  &  -557.90  &  -562.08  &  -564.84  &  -639.06  \\
2.0  &  -631.96  &  -635.85  &  -637.31  &  -711.29  \\
2.1  &  -680.64  &  -684.30  &  -684.95  &  -758.71  \\
2.2  &  -710.17  &  -713.63  &  -713.80  &  -787.37  \\
2.3  &  -725.38  &  -728.68  &  -728.57  &  -801.98  \\
2.35 &  -728.82  &  -732.05  &  -731.86  &  -805.19  \\
2.4  &  -730.02  &  -733.18  &  -732.93  &  -806.19  \\
2.5  &  -727.02  &  -730.05  &  -729.75  &  -802.89  \\
2.6  &  -718.65  &  -721.58  &  -721.28  &  -794.31  \\
2.7  &  -706.72  &  -709.54  &  -709.29  &  -782.22  \\
2.8  &  -692.64  &  -695.37  &  -695.19  &  -768.03  \\
2.9  &  -677.59  &  -680.23  &  -680.15  &  -752.93  \\
3.0  &  -662.64  &  -665.20  &  -665.25  &  -737.98  \\
3.2  &  -636.59  &  -638.95  &  -639.33  &  -711.89  \\
3.4  &  -615.74  &  -617.95  &  -618.53  &  -690.94  \\
3.6  &  -597.53  &  -599.65  &  -600.32  &  -672.66  \\
3.8  &  -581.54  &  -583.60  &  -584.32  &  -656.62  \\
4.0  &  -567.88  &  -569.91  &  -570.65  &  -642.92  \\
4.2  &  -556.62  &  -558.63  &  -559.38  &  -631.62  \\
4.4  &  -547.67  &  -549.67  &  -550.41  &  -622.64  \\
4.6  &  -540.83  &  -542.81  &  -543.54  &  -615.76  \\
4.8  &  -535.75  &  -537.73  &  -538.44  &  -610.67  \\
5.0  &  -532.08  &  -534.05  &  -534.75  &  -606.96  \\
\hline

\end{tabular}
\end{center}
\end{table}

\begin{figure}[t!]
 \includegraphics[width=0.45\textwidth]{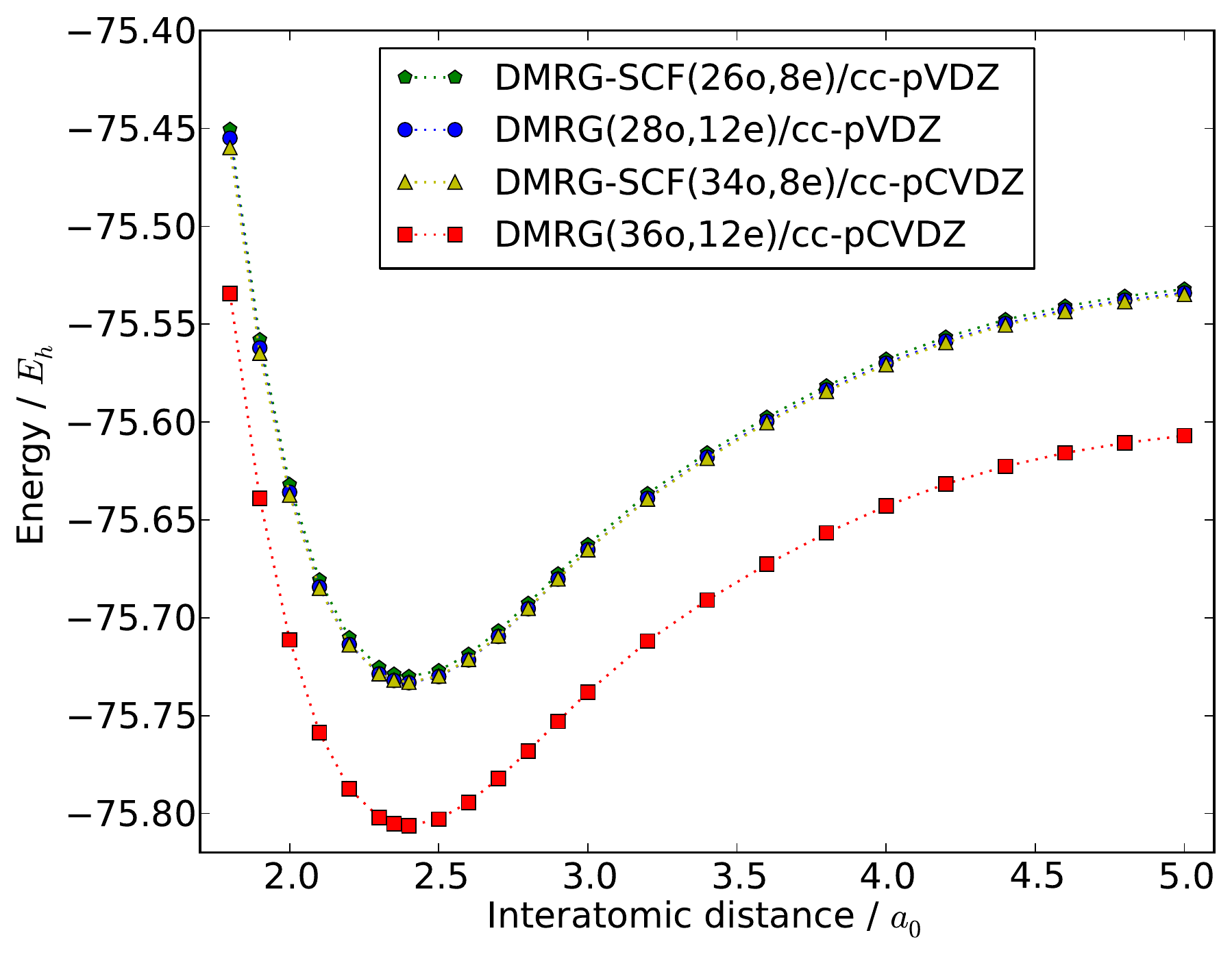}
 \caption{\label{CoreCorrAbsolute} Assessment of the importance of $1s$ core correlation. This effect is captured at the DMRG(36o, 12e, D$_{\mathsf{SU(2)}}$=2500)/cc-pCVDZ level of theory.}
\end{figure}
\begin{figure}[t!]
 \includegraphics[width=0.45\textwidth]{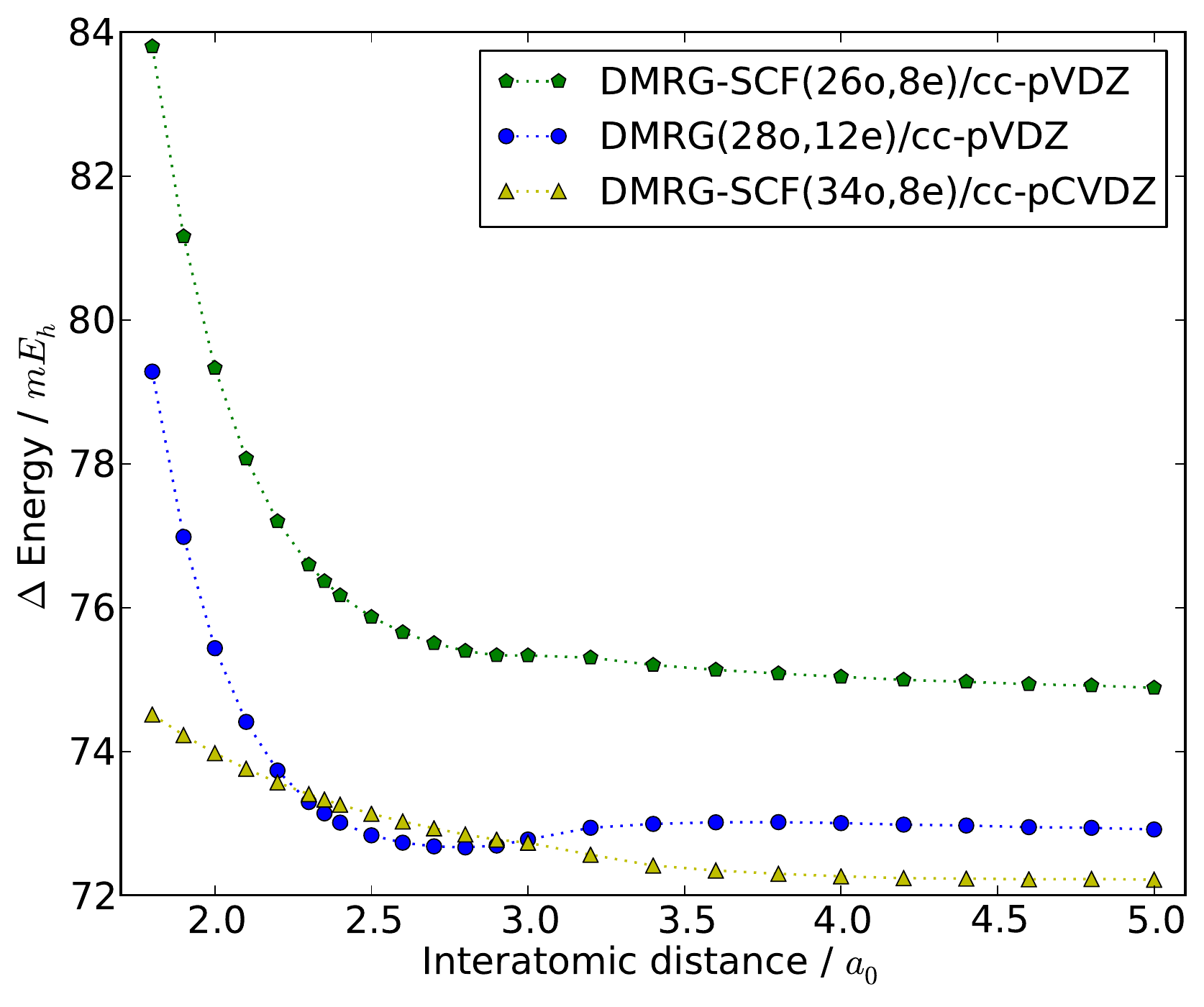}
 \caption{\label{CoreCorrRelative} Assessment of the importance of $1s$ core correlation. The relative energies with respect to the DMRG(36o, 12e, D$_{\mathsf{SU(2)}}$=2500)/cc-pCVDZ calculations are shown.}
\end{figure}

The extrapolated energies at the DMRG-SCF(26o, 8e, D$_{\mathsf{SU(2)}}$=2500)/cc-pVDZ, DMRG(28o, 12e, D$_{\mathsf{SU(2)}}$=2500)/cc-pVDZ, DMRG-SCF(34o, 8e, D$_{\mathsf{SU(2)}}$=2500)/cc-pCVDZ, and DMRG(36o, 12e, D$_{\mathsf{SU(2)}}$=2500)/cc-pCVDZ levels of theory are given in Tab. \ref{CoreCorrTable} and are shown in Fig. \ref{CoreCorrAbsolute}. The relative energies with respect to the DMRG(36o, 12e, D$_{\mathsf{SU(2)}}$=2500)/cc-pCVDZ calculations are shown in Fig. \ref{CoreCorrRelative}.

The $1s$ core correlation is only captured at the DMRG(36o, 12e, D$_{\mathsf{SU(2)}}$=2500)/cc-pCVDZ level of theory. Without the necessary orbital freedom, the $1s$ core correlation cannot be captured. The non-parallelity of the DMRG-SCF(34o, 8e, D$_{\mathsf{SU(2)}}$=2500)/cc-pCVDZ curve in Fig. \ref{CoreCorrRelative} is of the order of 2 $mE_h$, far below the error due to basis set incompleteness.

For small interatomic distances, the cc-pCVDZ curves show a different behaviour than the cc-pVDZ curves, as can be seen in Fig. \ref{CoreCorrRelative}. Extra basis set freedom is required to capture the more complicated core dynamics in the united atom limit. This can be understood as the transition from two light atoms, each with a doubly filled $1s$ orbital, to one single heavy atom, with several orbitals tightly packed around the nucleus.

\section{Summary} \label{summary}
In section \ref{intro}, we discussed how DMRG can be useful for ab initio quantum chemistry, and we gave an overview of DMRG-related methods. These methods can be divided into two categories: DMRG can play the role of a large active space FCI solver, or it can provide an approximate MPS wavefunction, on which excitations can be built.

The DMRG algorithm was introduced in section \ref{remarks}, where we discussed the use of complementary operators and how to overcome convergence difficulties. Both issues have to be addressed for DMRG to be an efficient and reliable approach for ab initio quantum chemistry.

With symmetry-adapted DMRG, a huge performance gain can be obtained both in computation time and memory. Section \ref{symm} introduced an MPS ansatz which is an exact eigenstate of the symmetry group of the Hamiltonian. The Wigner-Eckart theorem allows the introduction of a sparse block structure in this ansatz. For non-abelian groups, the Wigner-Eckart theorem also allows for data compression.

An overview of the high-level structure of CheMPS2 is given in section \ref{ourcode}. The required input for the \texttt{CheMPS2::DMRG} class and its output are discussed. A DMRG-SCF algorithm was implemented in \texttt{CheMPS2::CASSCF}. Section \ref{ourcode} should help new users to understand the provided tests, and to alter them to their own needs.

As an application, we have calculated the 12 lowest bond dissociation curves of the carbon dimer at the DMRG(28o, 12e, D$_{\mathsf{SU(2)}}$=2500)/cc-pVDZ level of theory. In addition, we assessed the contribution of $1s$ core correlation to the $X^1\Sigma_g^+$ bond dissociation curve of the carbon dimer by comparing calculations at the DMRG(36o, 12e, D$_{\mathsf{SU(2)}}$=2500)/cc-pCVDZ and DMRG-SCF(34o, 8e, D$_{\mathsf{SU(2)}}$=2500)/cc-pCVDZ levels of theory. These results were presented in section \ref{C2}. The low-lying bond dissociation curves of the carbon dimer were resolved with CheMPS2 to sub-$mE_h$ accuracy. The non-parallelity due to $1s$ core correlation is of the order of 2 $mE_h$ in the cc-pCVDZ basis.

In the future, we would like to incorporate the two-orbital mutual information $I_{p,q}$ \cite{Rissler2006519} in CheMPS2, as well as its gradient and hessian, to retrieve optimal orbitals and their corresponding ordering, as discussed in section \ref{ConvergenceSection}.

We are also working on an MPI implementation of CheMPS2, in which the product $\mathbf{H}^{eff} \mathbf{B}[i]$ is distributed over several processors. Each processor is then responsible for certain renormalized operators \cite{chan:3172}. Updated versions of CheMPS2 will be provided at its public git repository \cite{CheMPS2github}.

The oxo-Mn(salen) complex \cite{IvanicCite, SearsCite} is a great challenge for molecular electronic structure methods. We are currently performing large active space DMRG-SCF calculations with CheMPS2 to provide new insights in the relative order of the lowest singlet, triplet, and quintet states. Understanding the active space structure of this complex and several of its transition states will be of benefit for the experimentalists in our group \cite{C3CC44473B}.


\section*{Acknowledgements}
S.W. received a Ph.D. fellowship from the Research Foundation Flanders (FWO Vlaanderen). W.P. acknowledges support from a project funded by the Research Foundation Flanders (FWO Vlaanderen). P.W.A. acknowledges support from NSERC. This work was carried out using the Stevin Supercomputer Infrastructure at Ghent University, funded by Ghent University, the Hercules Foundation and the Flemish Government - department EWI.

\appendix

\section{Reduced tensors} \label{redtensors}

Note that during a sweep, we work with left-normalized tensors to the left and right-normalized tensors to the right of the current position. Consider the following renormalized partial Hamiltonian term in the graphical notation \cite{Schollwock201196}:
\begin{equation}
\vcenter{\hbox{
\includegraphics[height=0.12\textwidth]{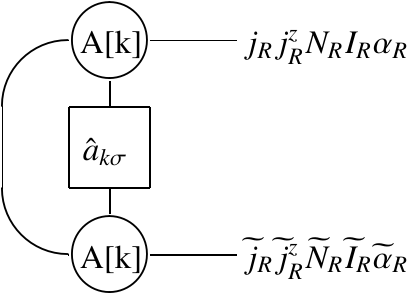} \label{example1}
}} \label{example1}
\end{equation}
With (\ref{tensordecomp}), it is easy to show that (\ref{example1}) can be written as
\begin{equation}
\delta_{N_R+1,\widetilde{N}_R} \delta_{I_R \otimes I_k, \widetilde{I}_R} \braket{j_R j_R^z \frac{1}{2} \sigma \mid \widetilde{j}_R \widetilde{j}_R^z} \vcenter{\hbox{
\includegraphics[height=0.12\textwidth]{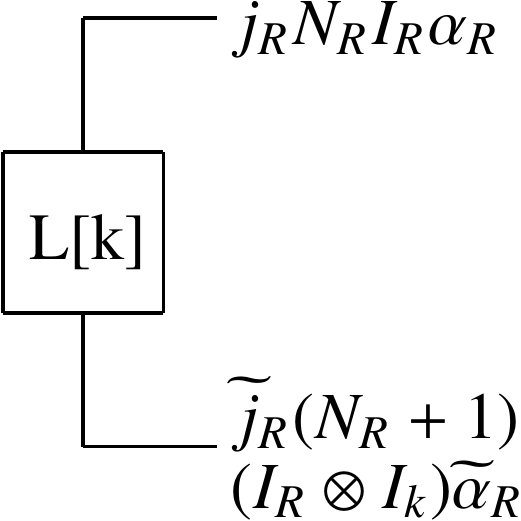} \label{againWE}
}}
\vspace{0.01\textwidth}
\end{equation}
with
\begin{eqnarray}
& \vcenter{\hbox{
\includegraphics[height=0.12\textwidth]{diagram2.pdf}
}}
= \sum\limits_{\alpha_L} \vcenter{\hbox{
\includegraphics[height=0.12\textwidth]{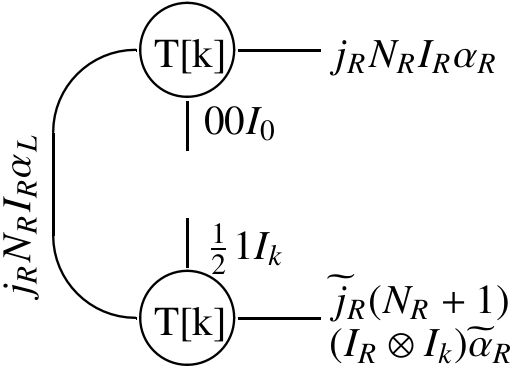}
}} \nonumber \\
& + (-1)^{\widetilde{j}_R - j_R + \frac{1}{2}} \sqrt{\frac{2 j_R + 1}{2 \widetilde{j}_R + 1}}\sum\limits_{\alpha_L} \vcenter{\hbox{
\includegraphics[height=0.12\textwidth]{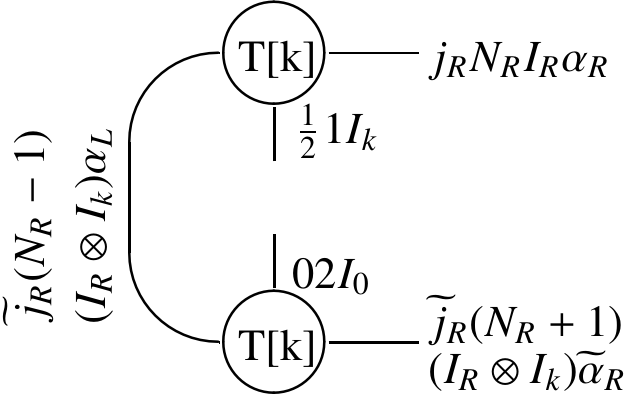}
}}
\end{eqnarray}
Eq. (\ref{example1}) can hence be factorized into Clebsch-Gordan coefficients and a reduced spin-$\frac{1}{2}$ $L$-tensor. The $L$-tensor has spin-$\frac{1}{2}$ because $\hat{a}_{k \sigma}$ is a spin-$\frac{1}{2}$ operator.

It is shown in Ref. \cite{wouters}, that for two second quantized operators acting on different sites, the renormalized operator can be decomposed into two terms: one with a spin-0 reduced tensor and one with a spin-1 reduced tensor. This follows from $\mathsf{SU(2)}$ representation theory: $\frac{1}{2} \otimes \frac{1}{2} \approx 0 \oplus 1$.





\begin{thebibliography}{100}
\expandafter\ifx\csname url\endcsname\relax
  \def\url#1{\texttt{#1}}\fi
\expandafter\ifx\csname urlprefix\endcsname\relax\def\urlprefix{URL }\fi
\expandafter\ifx\csname href\endcsname\relax
  \def\href#1#2{#2} \def\path#1{#1}\fi

\bibitem{BookHelgaker}
T.~Helgaker, P.~{J\o{}rgensen}, J.~Olsen, Molecular electronic-structure
  theory, 1st Edition, Wiley New-York, 2000.

\bibitem{whiteQC}
S.~R. White, R.~L. Martin, Ab initio quantum chemistry using the density matrix
  renormalization group, J. Chem. Phys. 110~(9) (1999) 4127--4130.
\newblock \href {http://dx.doi.org/10.1063/1.478295}
  {\path{doi:10.1063/1.478295}}.

\bibitem{PhysRevLett.69.2863}
S.~R. White, Density matrix formulation for quantum renormalization groups,
  Phys. Rev. Lett. 69 (1992) 2863--2866.
\newblock \href {http://dx.doi.org/10.1103/PhysRevLett.69.2863}
  {\path{doi:10.1103/PhysRevLett.69.2863}}.

\bibitem{PhysRevB.48.10345}
S.~R. White, Density-matrix algorithms for quantum renormalization groups,
  Phys. Rev. B 48 (1993) 10345--10356.
\newblock \href {http://dx.doi.org/10.1103/PhysRevB.48.10345}
  {\path{doi:10.1103/PhysRevB.48.10345}}.

\bibitem{PhysRevLett.75.3537}
S.~\"Ostlund, S.~Rommer, Thermodynamic limit of density matrix renormalization,
  Phys. Rev. Lett. 75 (1995) 3537--3540.
\newblock \href {http://dx.doi.org/10.1103/PhysRevLett.75.3537}
  {\path{doi:10.1103/PhysRevLett.75.3537}}.

\bibitem{PhysRevB.55.2164}
S.~Rommer, S.~\"Ostlund, Class of ansatz wave functions for one-dimensional
  spin systems and their relation to the density matrix renormalization group,
  Phys. Rev. B 55 (1997) 2164--2181.
\newblock \href {http://dx.doi.org/10.1103/PhysRevB.55.2164}
  {\path{doi:10.1103/PhysRevB.55.2164}}.

\bibitem{1742-5468-2007-08-P08024}
M.~B. Hastings, An area law for one-dimensional quantum systems, J. Stat. Mech.
  Theor. Exp. 08 (2007) P08024.
\newblock \href {http://dx.doi.org/10.1088/1742-5468/2007/08/P08024}
  {\path{doi:10.1088/1742-5468/2007/08/P08024}}.

\bibitem{PEPSverstraete}
F.~{Verstraete}, J.~I. {Cirac}, {Renormalization algorithms for Quantum-Many
  Body Systems in two and higher dimensions}, eprint
  arXiv:cond-mat/0407066.~\href {http://arxiv.org/abs/cond-mat/0407066}
  {\path{arXiv:cond-mat/0407066}}.

\bibitem{PhysRevLett.99.220405}
G.~Vidal, Entanglement renormalization, Phys. Rev. Lett. 99 (2007) 220405.
\newblock \href {http://dx.doi.org/10.1103/PhysRevLett.99.220405}
  {\path{doi:10.1103/PhysRevLett.99.220405}}.

\bibitem{PhysRevLett.104.190405}
F.~Verstraete, J.~I. Cirac, Continuous matrix product states for quantum
  fields, Phys. Rev. Lett. 104 (2010) 190405.
\newblock \href {http://dx.doi.org/10.1103/PhysRevLett.104.190405}
  {\path{doi:10.1103/PhysRevLett.104.190405}}.

\bibitem{QUA:QUA1}
S.~Daul, I.~Ciofini, C.~Daul, S.~R. White, Full-ci quantum chemistry using the
  density matrix renormalization group, Int. J. Quantum Chem. 79~(6) (2000)
  331--342.
\newblock \href
  {http://dx.doi.org/10.1002/1097-461X(2000)79:6<331::AID-QUA1>3.0.CO;2-Y}
  {\path{doi:10.1002/1097-461X(2000)79:6<331::AID-QUA1>3.0.CO;2-Y}}.

\bibitem{mitrushenkov:6815}
A.~O. Mitrushenkov, G.~Fano, F.~Ortolani, R.~Linguerri, P.~Palmieri, Quantum
  chemistry using the density matrix renormalization group, J. Chem. Phys.
  115~(15) (2001) 6815--6821.
\newblock \href {http://dx.doi.org/10.1063/1.1389475}
  {\path{doi:10.1063/1.1389475}}.

\bibitem{chan:4462}
G.~K.-L. Chan, M.~Head-Gordon, Highly correlated calculations with a polynomial
  cost algorithm: A study of the density matrix renormalization group, J. Chem.
  Phys. 116~(11) (2002) 4462--4476.
\newblock \href {http://dx.doi.org/10.1063/1.1449459}
  {\path{doi:10.1063/1.1449459}}.

\bibitem{PhysRevB.67.125114}
O.~Legeza, J.~R\"oder, B.~A. Hess, Controlling the accuracy of the
  density-matrix renormalization-group method: The dynamical block state
  selection approach, Phys. Rev. B 67 (2003) 125114.
\newblock \href {http://dx.doi.org/10.1103/PhysRevB.67.125114}
  {\path{doi:10.1103/PhysRevB.67.125114}}.

\bibitem{chan:8551}
G.~K.-L. Chan, M.~Head-Gordon, Exact solution (within a triple-zeta, double
  polarization basis set) of the electronic {S}chr\"odinger equation for water,
  J. Chem. Phys. 118~(19) (2003) 8551--8554.
\newblock \href {http://dx.doi.org/10.1063/1.1574318}
  {\path{doi:10.1063/1.1574318}}.

\bibitem{doi:10.1080/0026897031000155625}
O.~Legeza, J.~R\"oder, B.~A. Hess, {QC-DMRG} study of the ionic-neutral curve
  crossing of {LiF}, Mol. Phys. 101~(13) (2003) 2019--2028.
\newblock \href {http://dx.doi.org/10.1080/0026897031000155625}
  {\path{doi:10.1080/0026897031000155625}}.

\bibitem{mitrushenkov:4148}
A.~O. Mitrushenkov, R.~Linguerri, P.~Palmieri, G.~Fano, Quantum chemistry using
  the density matrix renormalization group {II}, J. Chem. Phys. 119~(8) (2003)
  4148--4158.
\newblock \href {http://dx.doi.org/10.1063/1.1593627}
  {\path{doi:10.1063/1.1593627}}.

\bibitem{PhysRevB.68.195116}
O.~Legeza, J.~S\'olyom, Optimizing the density-matrix renormalization group
  method using quantum information entropy, Phys. Rev. B 68 (2003) 195116.
\newblock \href {http://dx.doi.org/10.1103/PhysRevB.68.195116}
  {\path{doi:10.1103/PhysRevB.68.195116}}.

\bibitem{chan:3172}
G.~K.-L. Chan, An algorithm for large scale density matrix renormalization
  group calculations, J. Chem. Phys. 120~(7) (2004) 3172--3178.
\newblock \href {http://dx.doi.org/10.1063/1.1638734}
  {\path{doi:10.1063/1.1638734}}.

\bibitem{chan:6110}
G.~K.-L. Chan, M.~K\'{a}llay, J.~Gauss, State-of-the-art density matrix
  renormalization group and coupled cluster theory studies of the nitrogen
  binding curve, J. Chem. Phys. 121~(13) (2004) 6110--6116.
\newblock \href {http://dx.doi.org/10.1063/1.1783212}
  {\path{doi:10.1063/1.1783212}}.

\bibitem{PhysRevB.70.205118}
O.~Legeza, J.~S\'olyom, Quantum data compression, quantum information
  generation, and the density-matrix renormalization-group method, Phys. Rev. B
  70 (2004) 205118.
\newblock \href {http://dx.doi.org/10.1103/PhysRevB.70.205118}
  {\path{doi:10.1103/PhysRevB.70.205118}}.

\bibitem{moritz:024107}
G.~Moritz, B.~A. Hess, M.~Reiher, Convergence behavior of the density-matrix
  renormalization group algorithm for optimized orbital orderings, J. Chem.
  Phys. 122~(2) (2005) 024107.
\newblock \href {http://dx.doi.org/10.1063/1.1824891}
  {\path{doi:10.1063/1.1824891}}.

\bibitem{chan:204101}
G.~K.-L. Chan, T.~V. Voorhis, Density-matrix renormalization-group algorithms
  with nonorthogonal orbitals and non-hermitian operators, and applications to
  polyenes, J. Chem. Phys. 122~(20) (2005) 204101.
\newblock \href {http://dx.doi.org/10.1063/1.1899124}
  {\path{doi:10.1063/1.1899124}}.

\bibitem{moritz:184105}
G.~Moritz, A.~Wolf, M.~Reiher, Relativistic {DMRG} calculations on the curve
  crossing of cesium hydride, J. Chem. Phys. 123~(18) (2005) 184105.
\newblock \href {http://dx.doi.org/10.1063/1.2104447}
  {\path{doi:10.1063/1.2104447}}.

\bibitem{moritz:034103}
G.~Moritz, M.~Reiher, Construction of environment states in quantum-chemical
  density-matrix renormalization group calculations, J. Chem. Phys. 124~(3)
  (2006) 034103.
\newblock \href {http://dx.doi.org/10.1063/1.2139998}
  {\path{doi:10.1063/1.2139998}}.

\bibitem{hachmann:144101}
J.~Hachmann, W.~Cardoen, G.~K.-L. Chan, Multireference correlation in long
  molecules with the quadratic scaling density matrix renormalization group, J.
  Chem. Phys. 125~(14) (2006) 144101.
\newblock \href {http://dx.doi.org/10.1063/1.2345196}
  {\path{doi:10.1063/1.2345196}}.

\bibitem{Rissler2006519}
J.~Rissler, R.~M. Noack, S.~R. White, Measuring orbital interaction using
  quantum information theory, Chem. Phys. 323~(2–3) (2006) 519 -- 531.
\newblock \href {http://dx.doi.org/10.1016/j.chemphys.2005.10.018}
  {\path{doi:10.1016/j.chemphys.2005.10.018}}.

\bibitem{moritz:244109}
G.~Moritz, M.~Reiher, Decomposition of density matrix renormalization group
  states into a slater determinant basis, J. Chem. Phys. 126~(24) (2007)
  244109.
\newblock \href {http://dx.doi.org/10.1063/1.2741527}
  {\path{doi:10.1063/1.2741527}}.

\bibitem{dorando:084109}
J.~J. Dorando, J.~Hachmann, G.~K.-L. Chan, Targeted excited state algorithms,
  J. Chem. Phys. 127~(8) (2007) 084109.
\newblock \href {http://dx.doi.org/10.1063/1.2768360}
  {\path{doi:10.1063/1.2768360}}.

\bibitem{hachmann:134309}
J.~Hachmann, J.~J. Dorando, M.~Avil\'{e}s, G.~K.-L. Chan, The radical character
  of the acenes: A density matrix renormalization group study, J. Chem. Phys.
  127~(13) (2007) 134309.
\newblock \href {http://dx.doi.org/10.1063/1.2768362}
  {\path{doi:10.1063/1.2768362}}.

\bibitem{marti:014104}
K.~H. Marti, I.~M. Ond\'{\i}k, G.~Moritz, M.~Reiher, Density matrix
  renormalization group calculations on relative energies of transition metal
  complexes and clusters, J. Chem. Phys. 128~(1) (2008) 014104.
\newblock \href {http://dx.doi.org/10.1063/1.2805383}
  {\path{doi:10.1063/1.2805383}}.

\bibitem{zgid:014107}
D.~Zgid, M.~Nooijen, On the spin and symmetry adaptation of the density matrix
  renormalization group method, J. Chem. Phys. 128~(1) (2008) 014107.
\newblock \href {http://dx.doi.org/10.1063/1.2814150}
  {\path{doi:10.1063/1.2814150}}.

\bibitem{zgid:144115}
D.~Zgid, M.~Nooijen, Obtaining the two-body density matrix in the density
  matrix renormalization group method, J. Chem. Phys. 128~(14) (2008) 144115.
\newblock \href {http://dx.doi.org/10.1063/1.2883980}
  {\path{doi:10.1063/1.2883980}}.

\bibitem{zgid:144116}
D.~Zgid, M.~Nooijen, The density matrix renormalization group self-consistent
  field method: Orbital optimization with the density matrix renormalization
  group method in the active space, J. Chem. Phys. 128~(14) (2008) 144116.
\newblock \href {http://dx.doi.org/10.1063/1.2883981}
  {\path{doi:10.1063/1.2883981}}.

\bibitem{ghosh:144117}
D.~Ghosh, J.~Hachmann, T.~Yanai, G.~K.-L. Chan, Orbital optimization in the
  density matrix renormalization group, with applications to polyenes and
  beta-carotene, J. Chem. Phys. 128~(14) (2008) 144117.
\newblock \href {http://dx.doi.org/10.1063/1.2883976}
  {\path{doi:10.1063/1.2883976}}.

\bibitem{ChanB805292C}
G.~K.-L. Chan, Density matrix renormalisation group {L}agrangians, Phys. Chem.
  Chem. Phys. 10 (2008) 3454--3459.
\newblock \href {http://dx.doi.org/10.1039/B805292C}
  {\path{doi:10.1039/B805292C}}.

\bibitem{ChanQUA:QUA22099}
T.~Yanai, Y.~Kurashige, D.~Ghosh, G.~K.-L. Chan, Accelerating convergence in
  iterative solution for large-scale complete active space
  self-consistent-field calculations, Int. J. Quantum Chem. 109~(10) (2009)
  2178--2190.
\newblock \href {http://dx.doi.org/10.1002/qua.22099}
  {\path{doi:10.1002/qua.22099}}.

\bibitem{dorando:184111}
J.~J. Dorando, J.~Hachmann, G.~K.-L. Chan, Analytic response theory for the
  density matrix renormalization group, J. Chem. Phys. 130~(18) (2009) 184111.
\newblock \href {http://dx.doi.org/10.1063/1.3121422}
  {\path{doi:10.1063/1.3121422}}.

\bibitem{kurashige:234114}
Y.~Kurashige, T.~Yanai, High-performance ab initio density matrix
  renormalization group method: Applicability to large-scale multireference
  problems for metal compounds, J. Chem. Phys. 130~(23) (2009) 234114.
\newblock \href {http://dx.doi.org/10.1063/1.3152576}
  {\path{doi:10.1063/1.3152576}}.

\bibitem{yanai:024105}
T.~Yanai, Y.~Kurashige, E.~Neuscamman, G.~K.-L. Chan, Multireference quantum
  chemistry through a joint density matrix renormalization group and canonical
  transformation theory, J. Chem. Phys. 132~(2) (2010) 024105.
\newblock \href {http://dx.doi.org/10.1063/1.3275806}
  {\path{doi:10.1063/1.3275806}}.

\bibitem{neuscamman:024106}
E.~Neuscamman, T.~Yanai, G.~K.-L. Chan, Strongly contracted canonical
  transformation theory, J. Chem. Phys. 132~(2) (2010) 024106.
\newblock \href {http://dx.doi.org/10.1063/1.3274822}
  {\path{doi:10.1063/1.3274822}}.

\bibitem{doi:10.1080/00268971003657078}
K.~H. Marti, M.~Reiher, {DMRG} control using an automated {R}ichardson-type
  error protocol, Mol. Phys. 108~(3-4) (2010) 501--512.
\newblock \href {http://dx.doi.org/10.1080/00268971003657078}
  {\path{doi:10.1080/00268971003657078}}.

\bibitem{PhysRevB.81.235129}
H.-G. Luo, M.-P. Qin, T.~Xiang, Optimizing {H}artree-{F}ock orbitals by the
  density-matrix renormalization group, Phys. Rev. B 81 (2010) 235129.
\newblock \href {http://dx.doi.org/10.1103/PhysRevB.81.235129}
  {\path{doi:10.1103/PhysRevB.81.235129}}.

\bibitem{mizukami:091101}
W.~Mizukami, Y.~Kurashige, T.~Yanai, Communication: Novel quantum states of
  electron spins in polycarbenes from ab initio density matrix renormalization
  group calculations, J. Chem. Phys. 133~(9) (2010) 091101.
\newblock \href {http://dx.doi.org/10.1063/1.3476461}
  {\path{doi:10.1063/1.3476461}}.

\bibitem{PhysRevA.83.012508}
G.~Barcza, O.~Legeza, K.~H. Marti, M.~Reiher, Quantum-information analysis of
  electronic states of different molecular structures, Phys. Rev. A 83 (2011)
  012508.
\newblock \href {http://dx.doi.org/10.1103/PhysRevA.83.012508}
  {\path{doi:10.1103/PhysRevA.83.012508}}.

\bibitem{boguslawski:224101}
K.~Boguslawski, K.~H. Marti, M.~Reiher, Construction of {CASCI}-type wave
  functions for very large active spaces, J. Chem. Phys. 134~(22) (2011)
  224101.
\newblock \href {http://dx.doi.org/10.1063/1.3596482}
  {\path{doi:10.1063/1.3596482}}.

\bibitem{kurashige:094104}
Y.~Kurashige, T.~Yanai, Second-order perturbation theory with a density matrix
  renormalization group self-consistent field reference function: Theory and
  application to the study of chromium dimer, J. Chem. Phys. 135~(9) (2011)
  094104.
\newblock \href {http://dx.doi.org/10.1063/1.3629454}
  {\path{doi:10.1063/1.3629454}}.

\bibitem{QUA:QUA23173}
A.~O. Mitrushchenkov, G.~Fano, R.~Linguerri, P.~Palmieri, On the importance of
  orbital localization in {QC}-{DMRG} calculations, Int. J. Quantum Chem.
  112~(6) (2012) 1606--1619.
\newblock \href {http://dx.doi.org/10.1002/qua.23173}
  {\path{doi:10.1002/qua.23173}}.

\bibitem{sharma:124121}
S.~Sharma, G.~K.-L. Chan, Spin-adapted density matrix renormalization group
  algorithms for quantum chemistry, J. Chem. Phys. 136~(12) (2012) 124121.
\newblock \href {http://dx.doi.org/10.1063/1.3695642}
  {\path{doi:10.1063/1.3695642}}.

\bibitem{wouters}
S.~Wouters, P.~A. Limacher, D.~{Van Neck}, P.~W. Ayers, Longitudinal static
  optical properties of hydrogen chains: Finite field extrapolations of matrix
  product state calculations, J. Chem. Phys. 136~(13) (2012) 134110.
\newblock \href {http://dx.doi.org/10.1063/1.3700087}
  {\path{doi:10.1063/1.3700087}}.

\bibitem{doi:10.1021/ct300211j}
K.~Boguslawski, K.~H. Marti, O.~Legeza, M.~Reiher, Accurate ab initio spin
  densities, J. Chem. Theory Comput. 8~(6) (2012) 1970--1982.
\newblock \href {http://dx.doi.org/10.1021/ct300211j}
  {\path{doi:10.1021/ct300211j}}.

\bibitem{C2CP23767A}
T.~Yanai, Y.~Kurashige, E.~Neuscamman, G.~K.-L. Chan, Extended implementation
  of canonical transformation theory: parallelization and a new level-shifted
  condition, Phys. Chem. Chem. Phys. 14 (2012) 7809--7820.
\newblock \href {http://dx.doi.org/10.1039/C2CP23767A}
  {\path{doi:10.1039/C2CP23767A}}.

\bibitem{doi:10.1021/jz301319v}
K.~Boguslawski, P.~Tecmer, O.~Legeza, M.~Reiher, Entanglement measures for
  single- and multireference correlation effects, J. Phys. Chem. Lett. 3~(21)
  (2012) 3129--3135.
\newblock \href {http://dx.doi.org/10.1021/jz301319v}
  {\path{doi:10.1021/jz301319v}}.

\bibitem{doi:10.1021/ct3008974}
W.~Mizukami, Y.~Kurashige, T.~Yanai, More $\pi$ electrons make a difference:
  Emergence of many radicals on graphene nanoribbons studied by ab initio
  {DMRG} theory, J. Chem. Theory Comput. 9~(1) (2013) 401--407.
\newblock \href {http://dx.doi.org/10.1021/ct3008974}
  {\path{doi:10.1021/ct3008974}}.

\bibitem{doi:10.1021/ct400247p}
K.~Boguslawski, P.~Tecmer, G.~Barcza, �.~Legeza, M.~Reiher, Orbital
  entanglement in bond-formation processes, J. Chem. Theory Comput. 9~(7)
  (2013) 2959--2973.
\newblock \href {http://dx.doi.org/10.1021/ct400247p}
  {\path{doi:10.1021/ct400247p}}.

\bibitem{naturechem}
Y.~Kurashige, G.~K.-L. Chan, T.~Yanai, Entangled quantum electronic
  wavefunctions of the {M}n4{C}a{O}5 cluster in photosystem {II}, Nat. Chem. 5
  (2013) 660--666.
\newblock \href {http://dx.doi.org/10.1038/nchem.1677}
  {\path{doi:10.1038/nchem.1677}}.

\bibitem{ma:224105}
Y.~Ma, H.~Ma, Assessment of various natural orbitals as the basis of large
  active space density-matrix renormalization group calculations, J. Chem.
  Phys. 138~(22) (2013) 224105.
\newblock \href {http://dx.doi.org/10.1063/1.4809682}
  {\path{doi:10.1063/1.4809682}}.

\bibitem{saitow:044118}
M.~Saitow, Y.~Kurashige, T.~Yanai, Multireference configuration interaction
  theory using cumulant reconstruction with internal contraction of density
  matrix renormalization group wave function, J. Chem. Phys. 139~(4) (2013)
  044118.
\newblock \href {http://dx.doi.org/10.1063/1.4816627}
  {\path{doi:10.1063/1.4816627}}.

\bibitem{doi:10.1021/ct400707k}
F.~Liu, Y.~Kurashige, T.~Yanai, K.~Morokuma, Multireference ab initio density
  matrix renormalization group ({DMRG})-{CASSCF} and {DMRG}-{CASPT2} study on
  the photochromic ring opening of spiropyran, J. Chem. Theory Comput. 9~(10)
  (2013) 4462--4469.
\newblock \href {http://dx.doi.org/10.1021/ct400707k}
  {\path{doi:10.1021/ct400707k}}.

\bibitem{C3CP53975J}
P.~Tecmer, K.~Boguslawski, O.~Legeza, M.~Reiher, Unravelling the
  quantum-entanglement effect of noble gas coordination on the spin ground
  state of {CUO}, Phys. Chem. Chem. Phys. 16 (2014) 719--727.
\newblock \href {http://dx.doi.org/10.1039/C3CP53975J}
  {\path{doi:10.1039/C3CP53975J}}.

\bibitem{KnechtPaper}
S.~Knecht, O.~Legeza, M.~Reiher, {Four-Component Density Matrix Renormalization
  Group}, eprint arXiv:1312.0970.~\href {http://arxiv.org/abs/1312.0970}
  {\path{arXiv:1312.0970}}.

\bibitem{HallbergBook}
K.~Hallberg, Density matrix renormalization, in: D.~S\'en\'echal, A.-M.
  Tremblay, C.~Bourbonnais (Eds.), Theoretical Methods for Strongly Correlated
  Electrons, CRM Series in Mathematical Physics, Springer New York, 2004, pp.
  3--37.
\newblock \href {http://dx.doi.org/10.1007/0-387-21717-7_1}
  {\path{doi:10.1007/0-387-21717-7_1}}.

\bibitem{PhysRevB.85.035130}
B.~Pirvu, J.~Haegeman, F.~Verstraete, Matrix product state based algorithm for
  determining dispersion relations of quantum spin chains with periodic
  boundary conditions, Phys. Rev. B 85 (2012) 035130.
\newblock \href {http://dx.doi.org/10.1103/PhysRevB.85.035130}
  {\path{doi:10.1103/PhysRevB.85.035130}}.

\bibitem{PhysRevB.85.100408}
J.~Haegeman, B.~Pirvu, D.~J. Weir, J.~I. Cirac, T.~J. Osborne, H.~Verschelde,
  F.~Verstraete, Variational matrix product ansatz for dispersion relations,
  Phys. Rev. B 85 (2012) 100408.
\newblock \href {http://dx.doi.org/10.1103/PhysRevB.85.100408}
  {\path{doi:10.1103/PhysRevB.85.100408}}.

\bibitem{PhysRevB.88.075122}
S.~Wouters, N.~Nakatani, D.~Van~Neck, G.~K.-L. Chan, Thouless theorem for
  matrix product states and subsequent post density matrix renormalization
  group methods, Phys. Rev. B 88 (2013) 075122.
\newblock \href {http://dx.doi.org/10.1103/PhysRevB.88.075122}
  {\path{doi:10.1103/PhysRevB.88.075122}}.

\bibitem{PhysRevB.88.075133}
J.~Haegeman, T.~J. Osborne, F.~Verstraete, Post-matrix product state methods:
  To tangent space and beyond, Phys. Rev. B 88 (2013) 075133.
\newblock \href {http://dx.doi.org/10.1103/PhysRevB.88.075133}
  {\path{doi:10.1103/PhysRevB.88.075133}}.

\bibitem{2013arXiv1311.1646N}
N.~{Nakatani}, S.~{Wouters}, D.~{Van Neck}, G.-L. {Chan}, {Linear Response
  Theory for the Density Matrix Renormalization Group: Efficient Algorithms for
  Strongly Correlated Excited States}, eprint arXiv:1311.1646.~\href
  {http://arxiv.org/abs/1311.1646} {\path{arXiv:1311.1646}}.

\bibitem{PhysRevLett.107.070601}
J.~Haegeman, J.~I. Cirac, T.~J. Osborne, I.~Pi\ifmmode~\check{z}\else
  \v{z}\fi{}orn, H.~Verschelde, F.~Verstraete, Time-dependent variational
  principle for quantum lattices, Phys. Rev. Lett. 107 (2011) 070601.
\newblock \href {http://dx.doi.org/10.1103/PhysRevLett.107.070601}
  {\path{doi:10.1103/PhysRevLett.107.070601}}.

\bibitem{PhysRevB.82.205105}
V.~Murg, F.~Verstraete, O.~Legeza, R.~M. Noack, Simulating strongly correlated
  quantum systems with tree tensor networks, Phys. Rev. B 82 (2010) 205105.
\newblock \href {http://dx.doi.org/10.1103/PhysRevB.82.205105}
  {\path{doi:10.1103/PhysRevB.82.205105}}.

\bibitem{nakatani:134113}
N.~Nakatani, G.~K.-L. Chan, Efficient tree tensor network states ({TTNS}) for
  quantum chemistry: Generalizations of the density matrix renormalization
  group algorithm, J. Chem. Phys. 138~(13) (2013) 134113.
\newblock \href {http://dx.doi.org/10.1063/1.4798639}
  {\path{doi:10.1063/1.4798639}}.

\bibitem{1367-2630-12-10-103008}
K.~H. Marti, B.~Bauer, M.~Reiher, M.~Troyer, F.~Verstraete, Complete-graph
  tensor network states: a new fermionic wave function ansatz for molecules,
  New J. Phys. 12~(10) (2010) 103008.
\newblock \href {http://dx.doi.org/10.1088/1367-2630/12/10/103008}
  {\path{doi:10.1088/1367-2630/12/10/103008}}.

\bibitem{RevModPhys.84.1527}
P.~J. Mohr, B.~N. Taylor, D.~B. Newell, Codata recommended values of the
  fundamental physical constants: 2010, Rev. Mod. Phys. 84 (2012) 1527--1605.
\newblock \href {http://dx.doi.org/10.1103/RevModPhys.84.1527}
  {\path{doi:10.1103/RevModPhys.84.1527}}.

\bibitem{Davidson197587}
E.~R. Davidson, The iterative calculation of a few of the lowest eigenvalues
  and corresponding eigenvectors of large real-symmetric matrices, J. Comput.
  Phys. 17~(1) (1975) 87--94.
\newblock \href {http://dx.doi.org/10.1016/0021-9991(75)90065-0}
  {\path{doi:10.1016/0021-9991(75)90065-0}}.

\bibitem{PhysRevB.53.R10445}
T.~Xiang, Density-matrix renormalization-group method in momentum space, Phys.
  Rev. B 53 (1996) R10445--R10448.
\newblock \href {http://dx.doi.org/10.1103/PhysRevB.53.R10445}
  {\path{doi:10.1103/PhysRevB.53.R10445}}.

\bibitem{chanExtraPolWithAyers}
G.~K.-L. Chan, P.~W. Ayers, E.~S.~I. Croot, On the distribution of eigenvalues
  of grand canonical density matrices, J. Stat. Phys. 109~(1-2) (2002)
  289--299.
\newblock \href {http://dx.doi.org/10.1023/A:1019999930923}
  {\path{doi:10.1023/A:1019999930923}}.

\bibitem{PhysRevB.53.14349}
O.~Legeza, G.~F\'ath, Accuracy of the density-matrix renormalization-group
  method, Phys. Rev. B 53 (1996) 14349--14358.
\newblock \href {http://dx.doi.org/10.1103/PhysRevB.53.14349}
  {\path{doi:10.1103/PhysRevB.53.14349}}.

\bibitem{2013arXiv1307.1002V}
B.~{Verstichel}, W.~{Poelmans}, S.~{De Baerdemacker}, S.~{Wouters}, D.~{Van
  Neck}, {v2DM study of the 2D Hubbard model: Benchmark results with
  three-index conditions and extended cluster constraints}, eprint
  arXiv:1307.1002.~\href {http://arxiv.org/abs/1307.1002}
  {\path{arXiv:1307.1002}}.

\bibitem{2002EL57852M}
I.~P. {McCulloch}, M.~{Gul{\'a}csi}, {The non-Abelian density matrix
  renormalization group algorithm}, Europhys. Lett. 57 (2002) 852--858.
\newblock \href {http://dx.doi.org/10.1209/epl/i2002-00393-0}
  {\path{doi:10.1209/epl/i2002-00393-0}}.

\bibitem{2007JSMTE1014M}
I.~P. {McCulloch}, {From density-matrix renormalization group to matrix product
  states}, J. Stat. Mech. Theor. Exp. 10 (2007) P10014.
\newblock \href {http://dx.doi.org/10.1088/1742-5468/2007/10/P10014}
  {\path{doi:10.1088/1742-5468/2007/10/P10014}}.

\bibitem{2010NJPh12c3029S}
S.~{Singh}, H.-Q. {Zhou}, G.~{Vidal}, {Simulation of one-dimensional quantum
  systems with a global SU(2) symmetry}, New J. Phys. 12~(3) (2010) 033029.
\newblock \href {http://dx.doi.org/10.1088/1367-2630/12/3/033029}
  {\path{doi:10.1088/1367-2630/12/3/033029}}.

\bibitem{PhysRevA.82.050301}
S.~Singh, R.~N.~C. Pfeifer, G.~Vidal, Tensor network decompositions in the
  presence of a global symmetry, Phys. Rev. A 82 (2010) 050301.
\newblock \href {http://dx.doi.org/10.1103/PhysRevA.82.050301}
  {\path{doi:10.1103/PhysRevA.82.050301}}.

\bibitem{BookCornwell}
J.~F. Cornwell, Group theory in physics, 1st Edition, Vol. 1-2, Academic Press
  Inc. (London) Ltd., 1984.

\bibitem{BookDimitri}
W.~Dickhoff, D.~Van~Neck, Many-body theory exposed!, 2nd Edition, World
  Scientific, 2008.

\bibitem{GSLcitation}
{GNU} {S}cientific {L}ibrary 1.15, \url{http://www.gnu.org/software/gsl/}
  (2011).

\bibitem{CheMPS2github}
S.~Wouters, {CheMPS2}: a spin-adapted implementation of {DMRG} for ab initio
  quantum chemistry, \url{https://github.com/SebWouters/CheMPS2} (2013).

\bibitem{Psi4article}
J.~M. Turney, A.~C. Simmonett, R.~M. Parrish, E.~G. Hohenstein, F.~A.
  Evangelista, J.~T. Fermann, B.~J. Mintz, L.~A. Burns, J.~J. Wilke, M.~L.
  Abrams, N.~J. Russ, M.~L. Leininger, C.~L. Janssen, E.~T. Seidl, W.~D. Allen,
  H.~F. Schaefer, R.~A. King, E.~F. Valeev, C.~D. Sherrill, T.~D. Crawford,
  Psi4: an open-source ab initio electronic structure program, WIREs Comput.
  Mol. Sci. 2~(4) (2012) 556--565.
\newblock \href {http://dx.doi.org/10.1002/wcms.93}
  {\path{doi:10.1002/wcms.93}}.

\bibitem{LengsfieldPaper}
B.~H. Lengsfield, General second order {MCSCF} theory: A density matrix
  directed algorithm, J. Chem. Phys. 73~(1) (1980) 382--390.
\newblock \href {http://dx.doi.org/10.1063/1.439885}
  {\path{doi:10.1063/1.439885}}.

\bibitem{SiegbahnPaper}
P.~E.~M. Siegbahn, J.~Alml\"of, A.~Heiberg, B.~O. Roos, The complete active
  space {SCF} ({CASSCF}) method in a {N}ewton-{R}aphson formulation with
  application to the {HNO} molecule, J. Chem. Phys. 74~(4) (1981) 2384--2396.
\newblock \href {http://dx.doi.org/10.1063/1.441359}
  {\path{doi:10.1063/1.441359}}.

\bibitem{bondingConundrums}
P.~Su, J.~Wu, J.~Gu, W.~Wu, S.~Shaik, P.~C. Hiberty, Bonding conundrums in the
  {C2} molecule: A valence bond study, J. Chem. Theory Comput, 7~(1) (2011)
  121--130.
\newblock \href {http://dx.doi.org/10.1021/ct100577v}
  {\path{doi:10.1021/ct100577v}}.

\bibitem{ChargeShiftnaturechem}
S.~Shaik, D.~Danovich, W.~Wu, P.~C. Hiberty, Charge-shift bonding and its
  manifestations in chemistry, Nat. Chem. 1 (2009) 443--449.
\newblock \href {http://dx.doi.org/10.1038/nchem.327}
  {\path{doi:10.1038/nchem.327}}.

\bibitem{PhysRev.56.778}
R.~S. Mulliken, Note on electronic states of diatomic carbon, and the
  carbon-carbon bond, Phys. Rev. 56 (1939) 778--781.
\newblock \href {http://dx.doi.org/10.1103/PhysRev.56.778}
  {\path{doi:10.1103/PhysRev.56.778}}.

\bibitem{C2wu}
C.~J. Wu, E.~A. Carter, Ab initio thermochemistry for unsaturated {C2}
  hydrocarbons, J. Phys. Chem. 95~(21) (1991) 8352--8363.
\newblock \href {http://dx.doi.org/10.1021/j100174a058}
  {\path{doi:10.1021/j100174a058}}.

\bibitem{vonRaguSchleyer19936387}
P.~von Ragu\'eSchleyer, P.~Maslak, J.~Chandrasekhar, R.~S. Grev, Is a {CC}
  quadruple bond possible?, Tetrahedron Lett. 34~(40) (1993) 6387 -- 6390.
\newblock \href {http://dx.doi.org/10.1016/0040-4039(93)85052-X}
  {\path{doi:10.1016/0040-4039(93)85052-X}}.

\bibitem{weinhold2005valency}
F.~Weinhold, C.~R. Landis, Valency and bonding: a natural bond orbital
  donor-acceptor perspective, Cambridge University Press, 2005.

\bibitem{C2natchem}
S.~Shaik, D.~Danovich, W.~Wu, P.~Su, H.~S. Rzepa, P.~C. Hiberty, Quadruple
  bonding in {C2} and analogous eight-valence electron species, Nat. Chem. 4
  (2012) 195--200.
\newblock \href {http://dx.doi.org/10.1038/nchem.1263}
  {\path{doi:10.1038/nchem.1263}}.

\bibitem{ANIE201208206}
S.~Shaik, H.~S. Rzepa, R.~Hoffmann, One molecule, two atoms, three views, four
  bonds?, Angew. Chem. Int. Ed. 52~(10) (2013) 3020--3033.
\newblock \href {http://dx.doi.org/10.1002/anie.201208206}
  {\path{doi:10.1002/anie.201208206}}.

\bibitem{dunningC2}
L.~T. Xu, T.~H. Dunning, Insights into the perplexing nature of the bonding in
  {C2} from generalized valence bond calculations, J. Chem. Theory Comput. {In
  print}.
\newblock \href {http://dx.doi.org/10.1021/ct400867h}
  {\path{doi:10.1021/ct400867h}}.

\bibitem{TheoChemC2}
K.~A. Peterson, A.~K. Wilson, D.~E. Woon, T.~H. {Dunning Jr.}, Benchmark
  calculations with correlated molecular wave functions {XII}. {C}ore
  correlation effects on the homonuclear diatomic molecules {B2}-{F2}, Theor.
  Chem. Acc. 97~(1-4) (1997) 251--259.
\newblock \href {http://dx.doi.org/10.1007/s002140050259}
  {\path{doi:10.1007/s002140050259}}.

\bibitem{C2chinesen}
D.~Shi, X.~Zhang, J.~Sun, Z.~Zhu, {MRCI} study on spectroscopic and molecular
  properties of {$B^1\Delta_g$}, {$B'^1\Sigma^{+}_g$}, {$C^1\Pi_g$},
  {$D^1\Sigma^+_u$}, {$E^1\Sigma_g^+$} and {$1^1\Delta_u$} electronic states of
  the {C2} radical, Mol. Phys. 109~(11) (2011) 1453--1465.
\newblock \href {http://dx.doi.org/10.1080/00268976.2011.564593}
  {\path{doi:10.1080/00268976.2011.564593}}.

\bibitem{BoggioPasqua2000159}
M.~Boggio-Pasqua, A.~Voronin, P.~Halvick, J.-C. Rayez, Analytical
  representations of high level ab initio potential energy curves of the {C2}
  molecule, J. Mol. Struct. {THEOCHEM} 531~(1-3) (2000) 159 -- 167.
\newblock \href {http://dx.doi.org/10.1016/S0166-1280(00)00442-5}
  {\path{doi:10.1016/S0166-1280(00)00442-5}}.

\bibitem{abrams2004full}
M.~L. Abrams, C.~D. Sherrill, Full configuration interaction potential energy
  curves for the {$X^1\Sigma^+_g$}, {$B^1\Delta_g$}, and {$B'^1\Sigma_g^+$}
  states of {C2}: A challenge for approximate methods, J. Chem. Phys. 121~(19)
  (2004) 9211--9219.
\newblock \href {http://dx.doi.org/10.1063/1.1804498}
  {\path{doi:10.1063/1.1804498}}.

\bibitem{Varandas}
A.~J.~C. Varandas, Extrapolation to the complete-basis-set limit and the
  implications of avoided crossings: The {$X^1\Sigma^+_g$}, {$B^1\Delta_g$},
  and {$B'^1\Sigma_g^+$} states of {C2}, J. Chem. Phys. 129~(23) (2008) 234103.
\newblock \href {http://dx.doi.org/10.1063/1.3036115}
  {\path{doi:10.1063/1.3036115}}.

\bibitem{Kokkin}
D.~L. Kokkin, G.~B. Bacskay, T.~W. Schmidt, Oscillator strengths and radiative
  lifetimes for {C2}: {S}wan, {B}allik-{R}amsay, {P}hillips, and {$d3\Pi g
  \leftarrow c3\Sigma u+$} systems, J. Chem. Phys. 126~(8) (2007) 084302.
\newblock \href {http://dx.doi.org/10.1063/1.2436879}
  {\path{doi:10.1063/1.2436879}}.

\bibitem{JiangC2}
W.~Jiang, A.~K. Wilson, Multireference composite approaches for the accurate
  study of ground and excited electronic states: {C2}, {N2}, and {O2}, J. Chem.
  Phys. 134~(3) (2011) 034101.
\newblock \href {http://dx.doi.org/10.1063/1.3514031}
  {\path{doi:10.1063/1.3514031}}.

\bibitem{sherillC2comppaper}
C.~D. Sherrill, P.~Piecuch, The {$X^1\Sigma^+_g$}, {$B^1\Delta_g$} and
  {$B'^1\Sigma_g^+$} states of {C2}: A comparison of renormalized
  coupled-cluster and multireference methods with full configuration
  interaction benchmarks, J. Chem. Phys. 122~(12) (2005) 124104.
\newblock \href {http://dx.doi.org/10.1063/1.1867379}
  {\path{doi:10.1063/1.1867379}}.

\bibitem{useofAbrams2}
U.~S. Mahapatra, S.~Chattopadhyay, R.~K. Chaudhuri, Molecular applications of
  state-specific multireference perturbation theory to {HF}, {H2O}, {H2S},
  {C2}, and {N2} molecules, J. Chem. Phys. 129~(2) (2008) 024108.
\newblock \href {http://dx.doi.org/10.1063/1.2952666}
  {\path{doi:10.1063/1.2952666}}.

\bibitem{useofAbrams3}
W.~Purwanto, S.~Zhang, H.~Krakauer, Excited state calculations using phaseless
  auxiliary-field quantum {M}onte {C}arlo: Potential energy curves of low-lying
  {C2} singlet states, J. Chem. Phys. 130~(9) (2009) 094107.
\newblock \href {http://dx.doi.org/10.1063/1.3077920}
  {\path{doi:10.1063/1.3077920}}.

\bibitem{useofAbrams4}
G.~H. Booth, D.~Cleland, A.~J.~W. Thom, A.~Alavi, Breaking the carbon dimer:
  The challenges of multiple bond dissociation with full configuration
  interaction quantum {M}onte {C}arlo methods, J. Chem. Phys. 135~(8) (2011)
  084104.
\newblock \href {http://dx.doi.org/10.1063/1.3624383}
  {\path{doi:10.1063/1.3624383}}.

\bibitem{ccpcvdzreference}
D.~E. Woon, T.~H. Dunning, Gaussian basis sets for use in correlated molecular
  calculations. {V}. {C}ore - valence basis sets for boron through neon, J.
  Chem. Phys. 103~(11) (1995) 4572 -- 4585.
\newblock \href {http://dx.doi.org/10.1063/1.470645}
  {\path{doi:10.1063/1.470645}}.

\bibitem{IvanicCite}
J.~Ivanic, J.~R. Collins, S.~K. Burt, Theoretical study of the low lying
  electronic states of oxo{X}(salen) ({X} = {M}n, {M}n-, {F}e, and {C}r-)
  complexes, J. Phys. Chem. A 108~(12) (2004) 2314--2323.
\newblock \href {http://dx.doi.org/10.1021/jp031214g}
  {\path{doi:10.1021/jp031214g}}.

\bibitem{SearsCite}
J.~S. Sears, C.~D. Sherrill, The electronic structure of oxo-{M}n(salen):
  Single-reference and multireference approaches, J. Chem. Phys. 124~(14)
  (2006) 144314.
\newblock \href {http://dx.doi.org/10.1063/1.2187974}
  {\path{doi:10.1063/1.2187974}}.

\bibitem{C3CC44473B}
T.~Bogaerts, A.~{Van Yperen-De Deyne}, Y.-Y. Liu, F.~Lynen, V.~{Van
  Speybroeck}, P.~{Van Der Voort}, Mn-salen@{MIL}101({A}l): a heterogeneous{,}
  enantioselective catalyst synthesized using a {'}bottle around the ship{'}
  approach, Chem. Commun. 49 (2013) 8021--8023.
\newblock \href {http://dx.doi.org/10.1039/C3CC44473B}
  {\path{doi:10.1039/C3CC44473B}}.

\bibitem{Schollwock201196}
U.~Schollw\"ock, The density-matrix renormalization group in the age of matrix
  product states, Ann. Phys. 326~(1) (2011) 96 -- 192.
\newblock \href {http://dx.doi.org/10.1016/j.aop.2010.09.012}
  {\path{doi:10.1016/j.aop.2010.09.012}}.

\end{thebibliography}







\end{document}